\documentclass[journal=cmatex,manuscript=article]{achemso}

\usepackage[T1]{fontenc}
\usepackage[utf8]{inputenc}
\usepackage[english]{babel}

\usepackage[table]{xcolor}
\usepackage{chemformula}
\usepackage{hyperref}
\usepackage{cleveref}
\usepackage{siunitx}
\usepackage{xspace}
\usepackage{multirow}
\usepackage{hhline}

\newcommand{\cmcm}{$Cmcm$\xspace}
\newcommand{\cmcii}{$Cmc2_1$\xspace}
\newcommand{\piic}{$P2_1/c$\xspace}
\newcommand{\snbchx}{\ch{Sn2M(III)Ch2X3}\xspace}
\newcommand{\mixing}{\ch{Sn2M(III)_{1-{$a$}}M(III)'_{$a$}Ch_{2-{$b$}}Ch'_{$b$}X_{3-{$c$}}X'_{$c$}}\xspace}

\SectionNumbersOn
\crefname{section}{Sec.}{Secs.}
\crefname{equation}{Eq.}{Eqs.}
\crefname{figure}{Figure}{Figures}
\crefname{tabular}{Table}{Tables}
\crefname{table}{Table}{Tables}

\DeclareSIUnit\at{atom}
\DeclareSIUnit\angstrom{\text {Å}}

\author{Pascal Henkel}
\affiliation{Department of Applied Physics, Aalto University, P.O.Box 11100, FI-00076 AALTO, Finland}

\author{Jingrui Li}
\affiliation{Electronic Materials Research Laboratory, Key Laboratory of the Ministry of Education and International Center for Dielectric Research, School of Electronic Science and Engineering \&{} International Joint Laboratory for Micro/Nano Manufacturing and Measurement Technology, Xi'an Jiaotong University, Xi'an 710049, China}

\author{G. Krishnamurthy Grandhi}
\affiliation{Hybrid Solar Cells, Faculty of Engineering and Natural Sciences, Tampere University, P.O. Box 541, Tampere, FI, 33014, Finland}

\author{Paola Vivo}
\affiliation{Hybrid Solar Cells, Faculty of Engineering and Natural Sciences, Tampere University, P.O. Box 541, Tampere, FI, 33014, Finland}

\author{Patrick Rinke}
\affiliation{Department of Applied Physics, Aalto University, P.O.Box 11100, FI-00076 AALTO, Finland}
\email{patrick.rinke@aalto.fi}

\title{Screening Mixed-Metal \protect\ch{Sn2M(III)Ch2X3} Chalcohalides for Photovoltaic Applications}

\abbreviations{DFT, PBEsol, HSE06, SOC, PCE, XRD, LHP}

\begin{document}

\begin{tocentry}
\includegraphics[scale=1]{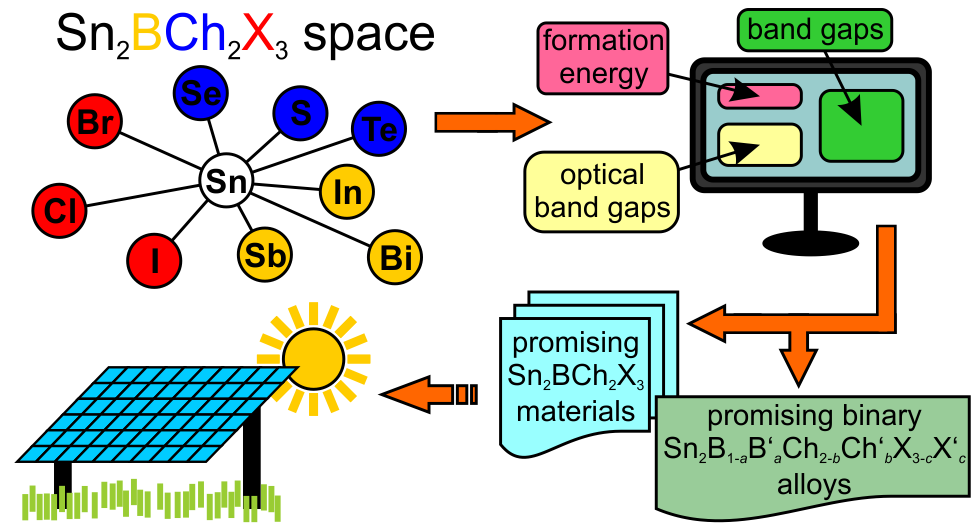}
\end{tocentry}

\begin{abstract}
Quaternary mixed-metal chalcohalides (\snbchx) are emerging as promising lead-free perovskite-inspired photovoltaic absorbers. Motivated by recent developments of a first \ch{Sn2SbS2I3}-based device, we used density functional theory to identify lead-free \snbchx materials that are structurally and energetically stable within \cmcm, \cmcii and \piic space groups and have a band gap in the range of \SIrange{0.7}{2.0}{\electronvolt} to cover out- and indoor photovoltaic applications. A total of 27 \snbchx materials were studied, including \ch{Sb}, \ch{Bi}, \ch{In} for \ch{M(III)}-site, \ch{S}, \ch{Se}, \ch{Te} for \ch{Ch}-site and \ch{Cl}, \ch{Br}, \ch{I} for \ch{X}-site. We identified 12 materials with a direct band gap that meet our requirements, namely: \ch{Sn2InS2Br3}, \ch{Sn2InS2I3}, \ch{Sn2InSe2Cl3}, \ch{Sn2InSe2Br3}, \ch{Sn2InTe2Br3}, \ch{Sn2InTe2Cl3}, \ch{Sn2SbS2I3}, \ch{Sn2SbSe2Cl3}, \ch{Sn2SbSe2I3}, \ch{Sn2SbTe2Cl3}, \ch{Sn2BiS2I3} and \ch{Sn2BiTe2Cl3}. A database scan reveals that 9 out of 12 are new compositions. For all 27 materials, \piic is the thermodynamically preferred structure, followed by \cmcii. In \cmcm and \cmcii mainly direct gaps occur, whereas mostly indirects in \piic. To open up the possibility of band gap tuning in the future, we identified 12 promising \mixing alloys which fulfill our requirements and additional 69 materials by combining direct and indirect band gap compounds.
\end{abstract}

\section{Introduction}
Photovoltaic technologies are instrumental for the transition from conventional to green and renewable energy production. To increase the power conversion efficiency while reducing costs and improving device longevity, new materials are continuously being explored. Lead halide perovskites (LHPs) have emerged as promising contenders due to their beneficial optoelectronic properties, defect-tolerance, and cost-effective solution processing. Yet, the toxicity of lead and the moderate long-term stability in air \cite{huang2021, huang2021b, ganose2017, nie2020b} impede the commercial viability of LHPs. \ch{Sn^{2+}}-based perovskites and \ch{Sb^{3+}}- and \ch{Bi^{3+}}-based perovskite-inspired absorbers are popular low-toxicity alternatives, but suffer from air-oxidation (\ch{Sn^{2+}} to \ch{Sn^{4+}})\cite{cao2021} and high defect densities.\cite{brandt2017, rondiya2021} Conversely, metal chalcogenides (e.g., \ch{Pb}-, \ch{Cd}-, \ch{Sb}-based)\cite{im2011, lee2009, choi2014} with tunable band gaps and high absorption cross-sections have enabled highly stable solar cells with modest efficiencies. Mixed-metal chalcohalides with \ch{M(II)2M(III)Ch2X3} stoichiometry (also referred to as \ch{A2BCh2X3})\cite{kavanagh2021}  are an emerging semiconductor family combining halide perovskite and metal chalcogenide building blocks.\cite{nie2020, kavanagh2021, sopiha2022} The \ch{Ch}-sites are occupied by the bivalent chalcogenide anions that form strong metal-chalcogen bonds with the metal cations. Mixed-metal chalcohalides may display the intriguing optoelectronic properties of LHPs, such as the dispersive valence and conduction bands, the high defect tolerance due to strong dielectric screening (owing to the presence of $ns^2$ lone pair cations) and the resultant low capture cross-sections of defects,\cite{huang2021, huang2021b, nie2020} with the promise to overcome the pressing stability challenges of LHPs.\cite{nie2020, nie2020b} In addition, the \ch{Sn}-\ch{Ch} bonding nature should prevent the \ch{Sn^{2+}} to \ch{Sn^{4+}} oxidation in mixed-metal chalcohalides. Also, experimental X-ray photoelectron spectroscopy measurements suggest that synthesizing \snbchx under reduction conditions partly suppresses \ch{Sn^{2+}} to \ch{Sn^{4+}} oxidation.\cite{nie2020} As a result, \snbchx films are stable even in  humid environments.\cite{nie2020, kavanagh2021} The photovoltaic potential of mixed-metal chalcohalides was recently demonstrated for a \ch{Sn2SbS2I3}-based single-junction solar cell that achieved  a power conversion efficiency (PCE) of \SI{4.04}{\percent}.\cite{nie2020} This is a promising start considering that the first perovskite solar cells only reached a PCE of \SI{3.8}{\percent} in 2009\cite{kojima2009} and now exceed  $\SI{25}{\percent}$.\cite{min2021}

Over the past four decades the \ch{M(II)2M(III)Ch2X3} material space – lead-free as well as the lead-based - has only  scarcely been explored (both theoretically and experimentally) and only a few compounds are known such as \ch{Sn2SbS2I3},\cite{olivier1980, nie2020, kavanagh2021, nicolson2023} \ch{Sn2SbSe2I3},\cite{ibanez1984} \ch{Sn2BiS2I3},\cite{islam2016} \ch{Pb2SbS2I3}\cite{dolgikh1985, doussier2007, roth2023} and \ch{Pb2BiS2I3}.\cite{islam2016, roth2023} X-ray diffraction (XRD) revealed that \ch{M(II)2M(III)Ch2X3} materials crystallize predominantly in an orthorhombic \cmcm space group.\cite{olivier1980, ibanez1984, dolgikh1985, doussier2007, islam2016, roth2023} For \ch{Sn2SbS2I3}, density functional theory (DFT) calculations demonstrated that this \cmcm structure has to be interpreted as an average over energetically more favorable, lower symmetry \cmcii configurations.\cite{kavanagh2021} XRD measurements by Doussier \textit{et al.} further found that \ch{Pb2SbS2I3} changes to a monoclinic \piic structure below \SI{100}{\kelvin}.\cite{doussier2007} This \piic structure was then shown to be lower in energy than the \cmcm and \cmcii phases for \ch{Sn2SbS2I3} by DFT.\cite{nicolson2023} In addition, for \ch{Sn2SbS2I3}, UV-vis absorption spectroscopy and DFT calculations revealed that the optical band gap lies below \SI{1.5}{\electronvolt}.\cite{starosta1990, nie2020, kavanagh2021} This limits its applications to single-junction solar cells as the band gap is close to the optimum value of harvesting solar radiation (\SI{1.3}{\electronvolt}).\cite{shockley1961, ruhle2016} In contrast, materials with a wide-band gap (\SIrange{1.6}{2.5}{\electronvolt}) will be of interest for emerging applications such as indoor and tandem photovoltaics.\cite{ganose2022}

Materials exploration can facilitate materials discovery for targeted properties and we apply it here to look for promising \snbchx materials. By now, many materials databases have been compiled \cite{himanen2019} and one could search them for materials that meet specified design criteria.\cite{seidu2019, shapera2018, guo2020} Materials or compounds that are not expected to be cataloged in databases can be explored by means of high-throughput computational or synthesis methods, often aided by machine-learning.\cite{schmidt2022, kim2018, kulik2022, schmidt2019, Ludwig2019,STEIN2022101053} Higher dimensional spaces offered by, e.g., quaternary or quinary materials still pose challenges, however, due to their sheer size and complexity. Such high-dimensional materials spaces have therefore only been explored partially. 

We here add an exploration of a \snbchx quarternary and quinary sub space for lead-free photovoltaic devices, which is chosen such that it can still be explored with DFT. For the \ch{M(III)}-site, we considered both \ch{Sb^{3+}} and \ch{Bi^{3+}} ($ns^2$ lone pair), and In(III) (with $ns^0$ ($d^{10}$) valence electron configuration). The chalcogen (\ch{Ch}) site is populated by \ch{S^{2-}}, \ch{Se^{2-}} or \ch{Te^{2-}}, and the halogen (\ch{X}) sites by \ch{Cl^{-}}, \ch{Br^{-}} or \ch{I^{-}}, which yields altogether 27 materials. We furthermore accounted for structural diversity by including the three reported space groups (\cmcm, \cmcii, and \piic).  

For each material and each phase, we carried out DFT calculations within the generalized gradient approximation to assess structural stability and with a hybrid functional to quantify the electronic structure. We then screened for thermodynamic stability and suitable band gap. Promising material candidates are cross-checked against a variety of databases and datasets to assess their novelty. Lastly, we investigated the potential to tailor the band gap by exploring different \mixing alloys.

The outline of this manuscript is as follows: In section II we describe our computational workflow including the screening criteria and the material cross-check. In Section III, we present and discuss our results for promising \snbchx materials, as well as for \mixing alloys. We conclude with a summary in Section IV.

\section{Computational Details}\label{sec:comp_details}
\subsection{Density-functional theory calculations}\label{sec:comp_details:dft}
We performed periodic, spin unpolarized DFT calculations with the all-electron, numeric atom-centered orbital code \textsc{FHI-AIMS} \cite{blum2009, havu2009, ren2012, levchenko2015, yu2018, ihrig2015}. For exchange and correlation (XC), we used he Perdew-Burke-Ernzerhof functional for solids (PBEsol) \cite{perdew2008, perdew2009}, which provides good agreement with experiment for the lattice constants of various halide perovskites at reasonable computational cost.\cite{yang2017, bokdam2017, li2023} The atomic structure  was relaxed with the Broyden-Fletcher-Goldfarb-Shanno algorithm and analytical stress tensor. \cite{knuth2015} Every structure was pre-optimized using \textit{light} real-space grid setting with a tier-1 basis set and was refined with \textit{tight} settings and a tier-2 basis set. We applied a Gaussian broadening of \SI{0.01}{\electronvolt} to the electronic occupations and relativistic effects were considered based on the zero-order regular approximation.\cite{blum2009, lenthe1993} The convergence threshold for the electronic self-consistency was set to \SI{1e-6}{\electronvolt}. The \snbchx structures were relaxed until all forces acting on the atoms were smaller than \SI{5e-3}{\electronvolt\per\angstrom}. Following the structural optimisations with PBEsol, we performed single point calculations for the band structure, the band gap, and the absorption spectra with the range-separated hybrid Heyd–Scuseria–Ernzerhof (HSE06) XC functional (with \SI{25}{\percent} exact exchange)\cite{heyd2003, heyd2006, krukau2006}, spin orbit coupling\cite{huhn2017} and also a tier-2 basis. The latter were calculated on the basis of the linear macroscopic dielectric tensor\cite{draxl2006} within the independent particle approximation.

We calculated each material in three crystal structures. For the \#{}63$/$\cmcm (see \cref{fig:a2bch2x3_spacegrp}a)) and \#{}36$/$\cmcii (see \cref{fig:a2bch2x3_spacegrp}b)) phases we used 16 atom unit cells and for the \#{}14$/$\piic (see \cref{fig:a2bch2x3_spacegrp}c)) phase a 32 atom unit cell. In  \cref{sec:comp_details:sn2bch2x3_struc} below, we describe the three crystal structures of \cmcm, \cmcii and \piic in more detail. A $\Gamma$-centered \textit{k}-point mesh was used for both PBEsol and HSE06 calculations. For the atomic structure relaxation with PBEsol, we used a $\Gamma$-centered $11 \times 11 \times 3$ $k$-point mesh for the \cmcm and \cmcii phases and a $6 \times 3 \times 5$ for \piic. The HSE band structure calculations we performed with $\Gamma$-centered $16 \times 16 \times 4$ for \cmcm and \cmcii and $9 \times 4 \times 8$ for \piic.

To estimate the thermodynamic stability (at \SI{0}{\kelvin}) we calculated the formation energy with respect to  elemental decomposition of an individual \snbchx material as 
\begin{equation}\label{eq:formationenergy}
E_\text{form}(\ch{Sn2M(III)Ch2X3}) = E_\text{tot}(\ch{Sn2M(III)Ch2X3}) - \sum_i x_i \mu_i.
\end{equation}
$E_\text{tot}(\ch{Sn2M(III)Ch2X3})$ denotes the (DFT-calculated) total energy of \snbchx, $\mu_i$ the chemical of the $i$th element and  $x_i$ the corresponding number of atoms. The upper limit for the  chemical potential is given by $\mu_i \leq E_\text{tot}(i\text{th element})$, i.e., the total energy per atom in the most stable phase of the $i$th element.\cite{vanderwalle2003} We used the following elemental compounds applying analogous computational parameters (with adjusted $\Gamma$-centered \textit{k}-point mesh): \ch{Sb} $R\overline{3}m$ (\#{}166), \ch{Bi} $R\overline{3}m$ (\#{}166), \ch{In} $I4/mmm$ (\#{}139), \ch{S} $Fddd$ (\#{}70), \ch{Se} $P2_1/c$ (\#{}14) and \ch{Te} $P3_121$ (\#{}152), as well as \ch{Cl}, \ch{Br} and \ch{I} in the gas phase as \ch{X_2}.

In the interest of open materials science \cite{himanen2019}, all relevant data is publicly available \href{https://Nomad-lab.eu/prod/v1/gui/dataset/id/nzZBqsV3TEGf7QYxc-5F3Q}{10.17172/NOMAD/2023.08.24-1}


\subsection{\protect\snbchx structures}\label{sec:comp_details:sn2bch2x3_struc}
\begin{figure}[ht]
\includegraphics[scale=0.275]{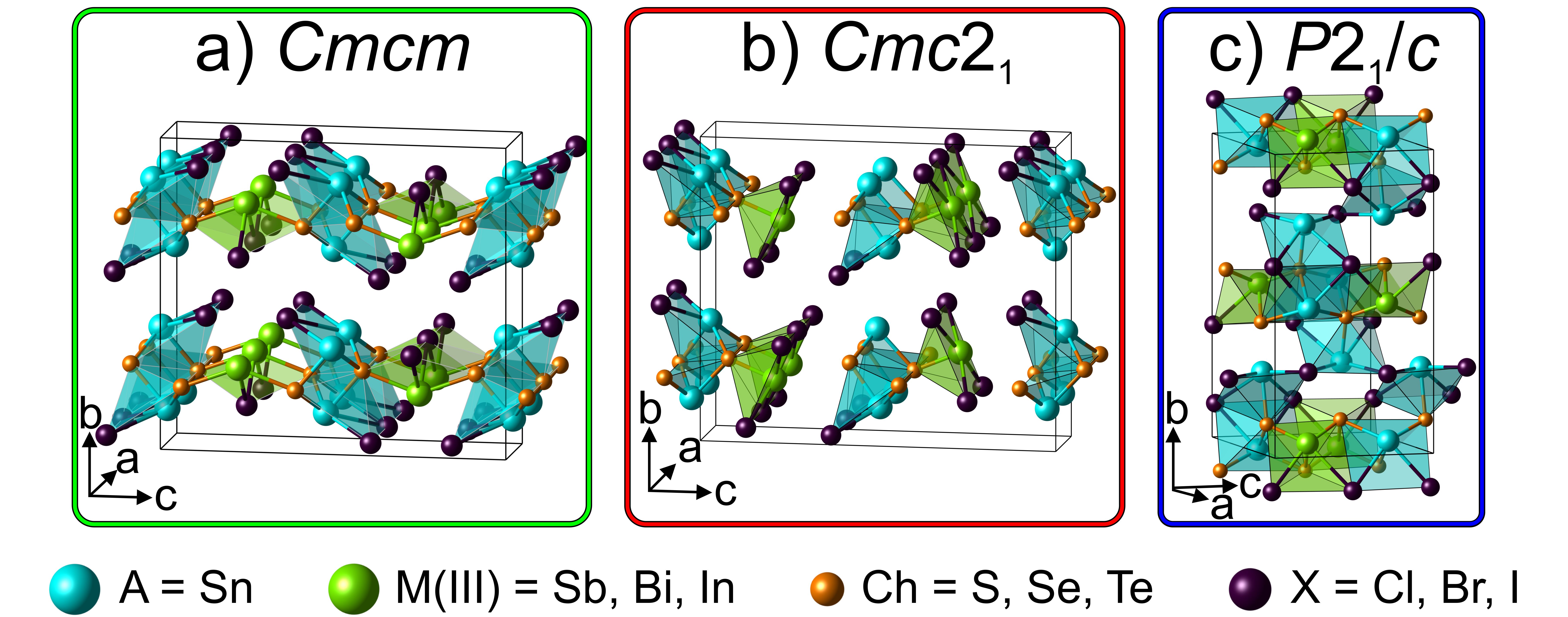}
\caption{Visualization of the \ch{Sn2M(III)Ch2X3} structure in different space groups; a) \cmcm, b) \cmcii and c) \piic for the example of \ch{Sn2SbS2I3}. The coordinating polyhedrons associated with the Sn/(A) positions are depicted in cyan and those associated with the \ch{M(III)} position in green. The coordinates of each space group were taken from the Materials Project\cite{materialsproject2013} and were visualized with VESTA.\cite{momma2008}}
\label{fig:a2bch2x3_spacegrp}
\end{figure}

\snbchx has been reported in three different space groups: \cmcm, \cmcii, and \piic (see \cref{fig:a2bch2x3_spacegrp}). In the \cmcm phase \ch{SnCh3X2} pyramids share edges and form periodically continued $[\ch{Sn2Ch2X2}]_n$ chains along the \textit{a} axis. The 1D $[\ch{Sn2Ch2X2}]_n$ chains in \snbchx \cmcm are connected by a \ch{M(III)X} unit along the \textit{c} axis to form a 2D \snbchx structure within the \textit{ac} plane. These 2D planes are not connected by chemical bonds along the \textit{b} axis (see \cref{fig:a2bch2x3_spacegrp}\,a)). In \cmcii (\cref{fig:a2bch2x3_spacegrp}\,b)), the \ch{M(III)X} element exclusively bonds to a single $[\ch{Sn2Ch2X2}]_n$ chain, forming a periodically linked $[\ch{Sn2M(III)Ch2X3}]_n$ chain along the \textit{a} axis, while the periodic connection along the \textit{c} axis is interrupted. Analogous to \cmcm, no linking is present along the \textit{b} axis. The asymmetric location of \ch{M(III)} between the chains (i.e., different to \cmcm) results in a reduction of every second square pyramidal \ch{SnCh2X3} polyhedron to an \ch{AX3} tetrahedron, while simultaneously forming an additional square pyramidal \ch{SnChX4} polyhedron. The \piic structure (\cref{fig:a2bch2x3_spacegrp}\,c)) is characterized by \ch{Sn4Ch4X8} units which are composed of four pyramids: \ch{SnChX4} - \ch{SnCh3X2} - \ch{SnCh3X2} - \ch{SnChX4}. \ch{SnChX4} and \ch{SnCh3X2} share faces and the two \ch{SnCh3X2} units edges. Along the \textit{c} axis, two neighbouring \ch{Sn4Ch4X6} units are linked by two \ch{M(III)} atoms, each forming a square pyramidal \ch{M(III)Ch3X2}. In addition, neighbouring \ch{Sn4Ch4X6} units share corners (\ch{X}) along the \textit{b} direction, thus forming a 2D \ch{Sn4M(III)2Ch4X6} structure within the \textit{bc} plane. 

We generated the initial structures of the individual \snbchx compounds from those of known materials listed in the Materials Project.\cite{materialsproject2013} For \cmcm and \cmcii, we adopted the \ch{Sn2SbS2I3} structures (ref. numbers: mp-561134 and mp-1219046) and for \piic, the \ch{Pb2SbS2I3} structure (ref. number: mp-578882). We then replaced any \ch{Pb} by \ch{Sn} and carried out corresponding replacements on the \ch{M(III)}, \ch{Ch} and \ch{X} positions to generate a total of 27 \snbchx compounds per space group. These were then geometry-optimized with PBEsol before we computed the electronic structure with HSE06+SOC.


\subsection{Screening criteria}
\label{sec:screening}

The aim of this study is to screen the mixed-metal chalcohalides \snbchx material space for compounds that are suitable for photovoltaic applications and have the potential to increase their PCE. We first consider formation energies  to estimate the materials' chemical stability. Chemical stability itself is hard to quantify, because it depends on the temperature, the chemical space, the synthesis process and the environmental conditions. To first order, we therefore approximate the chemical stability by the thermodynamic stability with respect to elementary decomposition computed at \SI{0}{\kelvin} as expressed in \cref{eq:formationenergy}. This allows us to draw first conclusions about the energetic preference of \snbchx compounds.

The band gap should be in the range of \SIrange{1.0}{1.8}{\electronvolt} for outdoor or \SIrange{1.5}{2.0}{\electronvolt} for indoor photovoltaics \cite{shockley1961, sutherland2020} according to the Shockley-Queisser limit. In our DFT calculations, we observe a difference of around \SI{0.3}{\electronvolt} between the fundamental and the optical band gap for some \snbchx materials (see discussion in \cref{sec:results_discuss:plain}). We therefore set a target range of \SIrange{0.7}{2.0}{\electronvolt} to encompass indoor and outdoor photovoltaic applications and incorporate the $\sim$\SI{0.3}{\electronvolt} shift.

\subsection{Novelty check}\label{sec:novelty_check}
To ascertain which materials in the \snbchx space have already been made or are known, we query a variety of materials databases. The \textsc{OPTIMADE} python API \cite{evans2021, andersen2021} provides an easy way to search a selection of databases with a single query. Our search includes the following 16 databases: Automatic FLOW for Materials Discovery (AFLOW),\cite{curtarolo2012, toher2018} Alexandria, Computational materials repository (CMR), Crystallography Open Database (COD),\cite{grazulis2009, grazulis2012} Cambridge Crystallographic Database (CCD),\cite{groom2016} Inorganic Crystal Structure Database (ICSD),\cite{bergerhoff1987,zagorac2019,allmann2007,belsky2002} Joint Automated Repository for Various Integrated Simulations (JARVIS),\cite{choudhary2020} Materials Cloud,\cite{ talirz2020,pizzi2016,huber2020} Materials Project,\cite{materialsproject2013} Materials Platform for Data Science (MPDS), Novel Materials Discovery Laboratory (NOMAD),\cite{draxl2018, ghiringhelli2017} Open Database of Xtals (odbx),\cite{evans2020} Open Materials Database (omdb),\cite{armiento2020} Open Quantum Materials Database (OQMD)\cite{saal2013} Theoretical Crystallography Open Database (TCOD),\cite{merkys2017} and 2D Materials Encyclopedia.\cite{zhou2019} In addition, we also included in our search the DCGAT-3 dataset by Schmidt \textit{et al.}, which was used to train their 3rd generation crystal-graph attention network and includes about 3.18 million crystalline compounds, including a large number of quaternary materials.\cite{schmidt2022}

\section{Results and discussion}\label{sec:results_discuss}

\subsection{Promising \protect\snbchx materials}\label{sec:results_discuss:plain}

\ch{Sn2SbS2I3} has already been studied with DFT by Kavanagh \textit{et al.} \cite{kavanagh2021} and Nicolson \textit{et al.} \cite{nicolson2023}, which provides us with an opportunity to validate our computational approach. Our PBEsol results for the energy differences between the \cmcii and \piic phases and the \cmcm phase are reported in \cref{tab:formationenergy_compare} and are in good agreement with the optB86b-vdW results of Kavanagh and Nicolson.

\begin{table}[ht]
\caption{Relative energies of both \cmcii and \piic phases of \ch{Sn2SbS2I3} with respect to the \cmcm phase (given in \si{\milli\electronvolt\per\at}). The energy of the reference \cmcm phase is set to $0$. Previous DFT results are included for comparison. Fundamental band gaps for \cmcm, \cmcii and \piic (given in \si{\electronvolt}). Indirect band gaps are marked with an *.}
\label{tab:formationenergy_compare}
\centering
\scalebox{.85}{
\begin{tabular}{l*{2}{p{1.25cm}}l|*{3}{p{1.25cm}}l} \hline
& $\Delta E_{\text{form}}^{}$ $(Cmc2_1^{})$ & $\Delta E_{\text{form}}^{}$ $(P2_1^{}/c)$ & Functional & $E_{g}^{}$ $(Cmcm^{})$ & $E_{g}^{}$ $(Cmc2_1^{})$ & $E_{g}^{}$ $(P2_1/c^{})$ & Functional \\\hline
Kavanagh \textit{et al.}\cite{kavanagh2021} & \num{-36} & -/- & optB86b-vdW & \num{1.02}$^*$ & \num{1.08} & -/- & HSE06+SOC\\
Nicolson \textit{et al.}\cite{nicolson2023} & \num{-11} & \num{-71} & optB86b-vdW & -/- & -/- & \num{1.78}$^*$ & HSE06+SOC \\
Our work & \num{-11} & \num{-69} & PBEsol & \num{0.947}$^*$ & \num{1.007} & \num{1.686}$^*$ & HSE06+SOC \\ \hline
\end{tabular}
}
\end{table}

Our band gap values calculated with HSE06 for \ch{Sn2SbS2I3} agree well with literature for all three phases, see \cref{tab:formationenergy_compare}. A direct gap is determined for \cmcii and indirect gaps for \cmcm and \piic. 

\begin{figure}[ht]
\includegraphics[scale=1.2125]{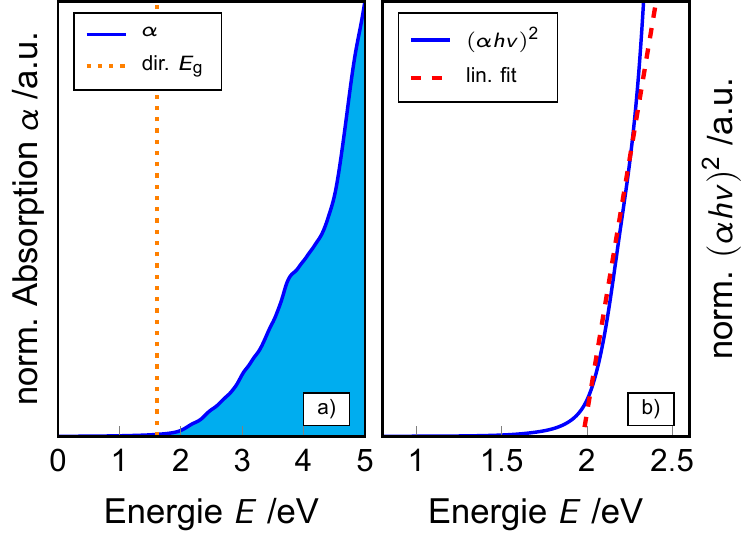}
\caption{Absorption spectrum normalized to \SI{5}{\electronvolt} (a) and resulting Tauc plot (b) (HSE06+SOC level of theory) for \ch{Sn2InS2Br3} within the \cmcm phase. The orange line in a) indicates the calculated fundamental band gap of \SI{1.617}{\electronvolt} and the red line in b) the best linear fit ($R^2 = \num{0.955}$) resulting in an optical band gap of $\sim\SI{1.977}{\electronvolt}$.}
\label{fig:example_absorption_tauc_plot}
\end{figure}

In addition, Kavanagh \textit{et al.} reported that the absorption edge is larger than the fundamental band gap for \ch{Sn2SbS2I3}. The authors proposed two reasons: I) a low density of states at the band edges caused by a low electronic degeneracy due to the low crystal symmetry (for \cmcm and \cmcii) and II) a weak transition dipole moment between the valence band maximum and the conduction band minimum due to the symmetry restrictions and a low spatial overlap.\cite{kavanagh2021} As a result, the absorption intensity is very low at the onset of absorption close to the fundamental band gap. It then rises slowly as the joint density of states increases. Appreciable absorption intensity is then only observed at some tenth of eV above the fundamental gap. To check if this behavior translates to other compounds in the \snbchx family, we calculated the absorption spectra for all 27 candidates. As an example, the absorption spectrum and the corresponding Tauc plot for \ch{Sn2InS2Br3} in the \cmcm phase is shown in \cref{fig:example_absorption_tauc_plot}. The intensity at the fundamental band gap is weak but not zero. It then increases rapidly at $E_\text{g} + \sim$\SI{0.4}, which is most evident in panel b). We define this transition point as ``optical gap''. From the Tauc plot, we deduce that the optical gap lies \SI{0.36}{\electronvolt} above the fundamental gap for \ch{Sn2InS2Br3}. The corresponding band structure is shown in Figure\,1 of the SI.

For all 27 \snbchx materials we find a larger optical gap, see Table\,1 in the SI. For compounds with a direct band gap, the optical band gap lies on average \SI{0.291}{\electronvolt} above the fundamental gap. This difference is space group dependent with $\sim\SI{0.376}{\electronvolt}$ for \cmcm and \cmcii and $\sim\SI{0.123}{\electronvolt}$ for \piic. For indirect band gap materials, the optical gap is also larger than the indirect fundamental gap ($E_\text{g, ind}$). However, if we compare to the lowest direct gap ($E_\text{g, dir}$), the difference to the optical gap is only \SI{0.049}{\electronvolt} on average and nearly identical for all space groups. This is caused by an increased spatial overlap between the valence band maximum and the conduction band. 
Therefore, our band gap criterion  considers not only the slightly different ranges for indoor and outdoor photovoltaics, but also the difference between the fundamental band gap and the absorption edge (see Section~\ref{sec:screening}). 

\begin{figure}[ht]
\includegraphics[scale=1.2125]{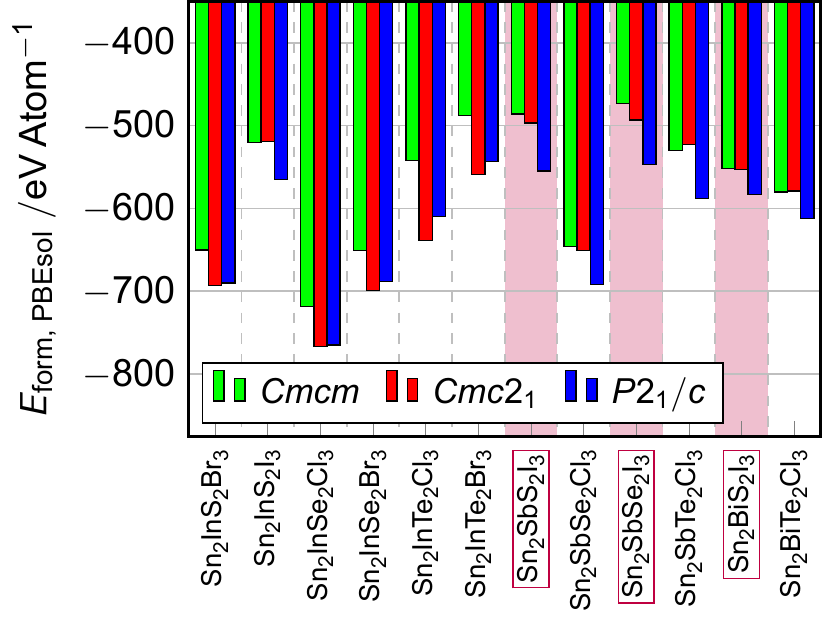}
\caption{Formation energy (calculated with PBEsol) of the 12 \ch{Sn2M(III)Ch2X3} materials of interest for \cmcm (green), \cmcii (red) and \piic (blue) space groups. Three structures - \ch{Sn2SbS2I3},\cite{olivier1980} \ch{Sn2SbSe2I3},\cite{ibanez1984} and \ch{Sn2BiS2I3}\cite{islam2016} - that are already known from the literature, are highlighted in purple.}
\label{fig:a2bch2x3_formationenergy}
\end{figure}

Our DFT calculations of the \snbchx materials space subsequently identified 12 compounds that satisfy the screening criteria of thermodynamic stability (negative formation energy at \SI{0}{\kelvin}) and fundamental direct band gap (in the range of \SIrange{0.7}{2.0}{\electronvolt}): \ch{Sn2InS2Br3}, \ch{Sn2InS2I3}, \ch{Sn2InSe2Cl3}, \ch{Sn2InSe2Br3}, \ch{Sn2InTe2Br3}, \ch{Sn2InTe2Cl3}, \ch{Sn2SbS2I3}, \ch{Sn2SbSe2Cl3}, \ch{Sn2SbSe2I3}, \ch{Sn2SbTe2Cl3}, \ch{Sn2BiS2I3} and \ch{Sn2BiTe2Cl3}. The formation energies of the 12 compounds in all three considered space groups are illustrated in \cref{fig:a2bch2x3_formationenergy}. A detailed overview of the formation energies of all 27 \snbchx compounds, as well as there lattice parameters, fundamental band and optical band gaps, is reported in Table\,1 of the SI.  

All of the 27 \snbchx structures are stable with respect to elemental decomposition (see Table\,1 of the SI) but not all fall into the desired band gap range. For the 12 candidates the most stable material is \ch{Sn2InSe2Cl3} (\SI{-718}, \SI{-767}, and \SI{-765}{\milli\electronvolt\per\at} for \cmcm, \cmcii, and \piic, respectively). The \ch{In}-based compounds are in most cases significantly more stable than their \ch{Sb}- and \ch{Bi}-based counterparts with the \cmcii phase being preferred. \ch{Sn2InS2I3},  in which the \piic structure is most stable, is an exception. For the \ch{Sb} or \ch{Bi} compounds, the \piic phase is the most stable. The formation energy decreases from \ch{Sn2M(III)Ch2Cl3} over \ch{Sn2M(III)Ch2Br3} towards \ch{Sn2M(III)Ch2I3}, which is natural as \ch{Cl} forms the strongest metal-halide bonds due to the highest electronegativity among the three halogen elements. 

An analogous behavior is obtained for 15 \snbchx compounds with an indirect fundamental gap (which we refer to as indirect \snbchx): \ch{Sn2InS2Cl3}, \ch{Sn2InSe2I3}, \ch{Sn2InTe2I3}, \ch{Sn2SbS2Cl3}, \ch{Sn2SbS2Br3}, \ch{Sn2SbSe2Br3}, \ch{Sn2SbTe2Br3}, \ch{Sn2SbTe2I3}, \ch{Sn2BiS2Cl3}, \ch{Sn2BiS2Br3}, \ch{Sn2BiSe2Cl3}, \ch{Sn2BiSe2Br3}, \ch{Sn2BiSe2I3}, \ch{Sn2BiTe2Br3} and \ch{Sn2BiTe2I3}. The lowest direct gap in these 15 materials is also within the screening range of \SIrange{0.7}{2.0}{\electronvolt}. Their formation energies are reported in Figure\,2 and Table\,1 in the SI.

\begin{figure}[ht]
\includegraphics[scale=1.2125]{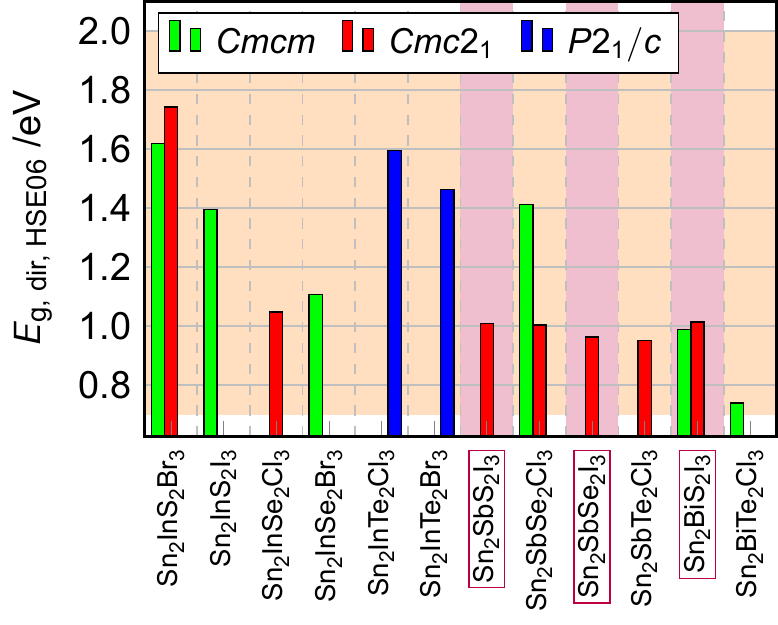}
\caption{Direct band gaps (HSE06+SOC level of theory) for 11 \ch{Sn2SbS2I3} materials of interest for \cmcm (green), \cmcii (red) and \piic (blue) space groups. Three structures - \ch{Sn2SbS2I3},\cite{olivier1980} \ch{Sn2SbSe2I3},\cite{ibanez1984} and \ch{Sn2BiS2I3}\cite{islam2016} - that are already known from the literature, are highlighted in purple.}
\label{fig:sn2bch2x3_fundbandgap_range}
\end{figure}

Our band gap results of the 27 \snbchx materials reveal that the space group influences the nature of the gap. For example, for \ch{Sn2InS2Br3} the \cmcm and \cmcii structures have a direct band gap, but not \piic. However, replacing \ch{S} with \ch{Se} leads to a direct band gap for the \cmcm structure and an indirect band gap for the other two structures. Overall, it turns out that all 12 identified materials have at least one space group with an indirect band gap. In \cref{fig:sn2bch2x3_fundbandgap_range} the direct band gaps for the 12 \snbchx materials are shown.

The \cmcm and \cmcii phases tend to have direct band gaps, whereas \piic structures exhibit indirect gaps. As a result, only two \piic compounds (\ch{Sn2InTe2Cl3} and \ch{Sn2InTe2Br3}) fall into our targeted band gap region. Also the band gap values depend on the space-group. Gaps of \cmcii structures lie around \SI{1}{\electronvolt} except for \ch{Sn2InS2Br3}. In contrast, \cmcm phases have larger band gaps $\sim$\SIrange{1.4}{1.5}{\electronvolt} with the exception of \ch{Sn2BiS2I3} and \ch{Sn2InSe2Br3}. In general,  direct band gaps are largest for \piic and smallest for \cmcii with \cmcm in between (see also \cref{tab:fundamental_bandgaps}). A similar picture emerges for the lowest direct band gaps in indirect gap materials (see \cref{tab:fundamental_bandgaps} and Figure\,3 of the SI).

\begin{table}[ht]
\caption{Fundamental and optical band gaps ($E_\text{g, opt}$) [given in \si{\electronvolt}] of \snbchx which have been identified as interesting. For indirect compounds the band gap of the lowest direct transition ($E_\text{g, dir}$) as well as of the lowest indirect transition ($E_\text{g, ind}$) are provided. \snbchx materials with a direct fundamental bang gap are colored in orange and these with an indirect one in gray.}
\label{tab:fundamental_bandgaps}
\centering
\scalebox{.7}{
\begin{tabular}{l|l|*{14}{p{1cm}}}
 & & \multicolumn{1}{c}{\rotatebox{90}{\ch{Sn2InS2Cl3}}} & \multicolumn{1}{c}{\rotatebox{90}{\ch{Sn2InS2Br3}}} & \multicolumn{1}{c}{\rotatebox{90}{\ch{Sn2InS2I3}}} & \multicolumn{1}{c}{\rotatebox{90}{\ch{Sn2InSe2Cl3}}} & \multicolumn{1}{c}{\rotatebox{90}{\ch{Sn2InSe2Br3}}} & \multicolumn{1}{c}{\rotatebox{90}{\ch{Sn2InSe2I3}}} & \multicolumn{1}{c}{\rotatebox{90}{\ch{Sn2InTe2Cl3}}} & \multicolumn{1}{c}{\rotatebox{90}{\ch{Sn2InTe2Br3}}} & \multicolumn{1}{c}{\rotatebox{90}{\ch{Sn2InTe2I3}}} & \multicolumn{1}{c}{\rotatebox{90}{\ch{Sn2SbS2Cl3}}} & \multicolumn{1}{c}{\rotatebox{90}{\ch{Sn2SbS2Br3}}} & \multicolumn{1}{c}{\rotatebox{90}{\ch{Sn2SbS2I3}}} & \multicolumn{1}{c}{\rotatebox{90}{\ch{Sn2SbSe2Cl3}}} & \multicolumn{1}{c}{\rotatebox{90}{\ch{Sn2SbSe2Br3}}} \\\hline
\multirow{3}{*}{\cmcm} & $E_\text{g, dir}$ & \cellcolor{gray!50}{\num{1.305}} & \cellcolor{orange!50}{\num{1.617}} & \cellcolor{orange!50}{\num{1.395}} & -/- & \cellcolor{orange!50}{\num{1.105}} & \cellcolor{gray!50}{\num{1.369}} & -/- & -/- & \cellcolor{gray!50}{\num{1.089}} & \cellcolor{gray!50}{\num{1.679}} & \cellcolor{gray!50}{\num{1.474}} & \cellcolor{gray!50}{\num{0.997}} & \cellcolor{orange!50}{\num{1.411}} & \cellcolor{gray!50}{\num{1.412}} \\
& $E_\text{g, ind}$ & \cellcolor{gray!50}{\num{1.251}} & \cellcolor{orange!50}{-/-} & \cellcolor{orange!50}{-/-} & -/- & \cellcolor{orange!50}{-/-} & \cellcolor{gray!50}{\num{1.299}} & -/- & -/- & \cellcolor{gray!50}{\num{0.994}} & \cellcolor{gray!50}{\num{1.611}} & \cellcolor{gray!50}{\num{0.883}} & \cellcolor{gray!50}{\num{0.947}} & \cellcolor{orange!50}{-/-} & \cellcolor{gray!50}{\num{1.330}} \\
& $E_\text{g, opt}$ & \cellcolor{gray!50}{\num{1.491}} & \cellcolor{orange!50}{\num{1.977}} & \cellcolor{orange!50}{\num{1.934}} & -/- & \cellcolor{orange!50}{\num{1.477}} & \cellcolor{gray!50}{\num{1.397}} & -/- & -/- & \cellcolor{gray!50}{\num{1.126}} & \cellcolor{gray!50}{\num{1.773}} & \cellcolor{gray!50}{\num{1.527}} & \cellcolor{gray!50}{\num{1.547}} & \cellcolor{orange!50}{\num{1.816}} & \cellcolor{gray!50}{\num{1.504}}  \\\hline
\multirow{3}{*}{\cmcii} & $E_\text{g, dir}$ & \cellcolor{gray!50}{\num{1.834}} & \cellcolor{orange!50}{\num{1.742}} & \cellcolor{gray!50}{\num{1.12}} & \cellcolor{orange!50}{\num{1.046}} & \cellcolor{gray!50}{\num{1.373}} & -/- & \cellcolor{gray!50}{\num{0.915}} & -/- & \cellcolor{gray!50}{\num{0.867}} & \cellcolor{gray!50}{\num{1.242}} & \cellcolor{gray!50}{\num{1.248}} & \cellcolor{orange!50}{\num{1.007}} & \cellcolor{orange!50}{\num{1.002}} & \cellcolor{gray!50}{\num{0.941}} \\
& $E_\text{g, ind}$ & \cellcolor{gray!50}{\num{1.692}} & \cellcolor{orange!50}{-/-} & \cellcolor{gray!50}{\num{0.598}} & \cellcolor{orange!50}{-/-} & \cellcolor{gray!50}{\num{1.338}} & -/- & \cellcolor{gray!50}{\num{0.804}} & -/- & \cellcolor{gray!50}{\num{0.802}} & \cellcolor{gray!50}{\num{1.056}} & \cellcolor{gray!50}{\num{1.055}} & \cellcolor{orange!50}{-/-} & \cellcolor{orange!50}{-/-} & \cellcolor{gray!50}{\num{0.891}} \\
& $E_\text{g, opt}$ & \cellcolor{gray!50}{\num{1.887}} & \cellcolor{orange!50}{\num{1.783}} & \cellcolor{gray!50}{\num{1.187}} & \cellcolor{orange!50}{\num{1.684}} & \cellcolor{gray!50}{\num{1.396}} & -/- & \cellcolor{gray!50}{\num{1.011}} & -/- & \cellcolor{gray!50}{\num{0.964}} & \cellcolor{gray!50}{\num{1.288}} & \cellcolor{gray!50}{\num{1.344}} & \cellcolor{orange!50}{\num{1.506}} & \cellcolor{orange!50}{\num{1.261}} & \cellcolor{gray!50}{\num{1.044}} \\\hline
\multirow{3}{*}{\piic} & $E_\text{g, dir}$ & -/- & \cellcolor{gray!50}{\num{1.985}} & -/- & -/- & \cellcolor{gray!50}{\num{1.843}} & \cellcolor{gray!50}{\num{1.947}} & \cellcolor{orange!50}{\num{1.594}} & \cellcolor{orange!50}{\num{1.461}} & \cellcolor{gray!50}{\num{1.109}} & -/- & -/- & \cellcolor{gray!50}{\num{1.762}} & -/- & \cellcolor{gray!50}{\num{2.028}} \\
& $E_\text{g, ind}$ & -/- & \cellcolor{gray!50}{\num{1.616}} & -/- & -/- & \cellcolor{gray!50}{\num{1.488}} & \cellcolor{gray!50}{\num{1.785}} & \cellcolor{orange!50}{-/-} & \cellcolor{orange!50}{-/-} & \cellcolor{gray!50}{\num{1.083}} & -/- & -/- & \cellcolor{gray!50}{\num{1.686}} & -/- & \cellcolor{gray!50}{\num{1.642}} \\
& $E_\text{g, opt}$ & -/- & \cellcolor{gray!50}{\num{1.996}} & -/- & -/- & \cellcolor{gray!50}{\num{1.855}} & \cellcolor{gray!50}{\num{1.960}} & \cellcolor{orange!50}{\num{1.713}} & \cellcolor{orange!50}{\num{1.592}} & \cellcolor{gray!50}{\num{1.201}} & -/- & -/- & \cellcolor{gray!50}{\num{1.771}} & -/- & \cellcolor{gray!50}{\num{2.100}} \\\hhline{=|=|==============}
& &  \multicolumn{1}{c}{\rotatebox{90}{\parbox{2.5cm}{\ch{Sn2SbSe2I3}}}} & \multicolumn{1}{c}{\rotatebox{90}{\ch{Sn2SbTe2Cl3}}} & \multicolumn{1}{c}{\rotatebox{90}{\ch{Sn2SbTe2Br3}}} & \multicolumn{1}{c}{\rotatebox{90}{\ch{Sn2SbTe2I3}}} & \multicolumn{1}{c}{\rotatebox{90}{\ch{Sn2BiS2Cl3}}} & \multicolumn{1}{c}{\rotatebox{90}{\ch{Sn2BiS2Br3}}} & \multicolumn{1}{c}{\rotatebox{90}{\ch{Sn2BiS2I3}}} & \multicolumn{1}{c}{\rotatebox{90}{\ch{Sn2BiSe2Cl3}}} & \multicolumn{1}{c}{\rotatebox{90}{\ch{Sn2BiSe2Br3}}} & \multicolumn{1}{c}{\rotatebox{90}{\ch{Sn2BiSe2I3}}}  & \multicolumn{1}{c}{\rotatebox{90}{\ch{Sn2BiTe2Cl3}}} & \multicolumn{1}{c}{\rotatebox{90}{\ch{Sn2BiTe2Br3}}} & \multicolumn{1}{c}{\rotatebox{90}{\ch{Sn2BiTe2I3}}} & \\\hline
\multirow{3}{*}{\cmcm} & $E_\text{g, dir}$ & \cellcolor{gray!50}{\num{1.057}} & \cellcolor{gray!50}{\num{0.943}} & -/- & \cellcolor{gray!50}{\num{0.786}} & \cellcolor{gray!50}{\num{1.535}} & \cellcolor{gray!50}{\num{1.343}} & \cellcolor{orange!50}{\num{0.988}} & \cellcolor{gray!50}{\num{1.104}} & \cellcolor{gray!50}{\num{0.965}} & \cellcolor{gray!50}{\num{1.048}} & \cellcolor{orange!50}{\num{0.738}} & \cellcolor{gray!50}{\num{0.852}} & -/- \\
& $E_\text{g, ind}$ & \cellcolor{gray!50}{\num{0.800}} & \cellcolor{gray!50}{\num{1.874}} & -/- & \cellcolor{gray!50}{\num{0.465}} & \cellcolor{gray!50}{\num{0.986}} & \cellcolor{gray!50}{\num{0.956}} & \cellcolor{orange!50}{-/-} & \cellcolor{gray!50}{\num{1.096}} & \cellcolor{gray!50}{\num{0.666}} & \cellcolor{gray!50}{\num{0.769}} & \cellcolor{orange!50}{-/-} & \cellcolor{gray!50}{\num{0.715}} & -/- \\
& $E_\text{g, opt}$ & \cellcolor{gray!50}{\num{1.056}} & \cellcolor{gray!50}{\num{1.997}} & -/- & \cellcolor{gray!50}{\num{0.795}} & \cellcolor{gray!50}{\num{1.548}} & \cellcolor{gray!50}{\num{1.434}} & \cellcolor{orange!50}{\num{1.390}} & \cellcolor{gray!50}{\num{1.124}} & \cellcolor{gray!50}{\num{1.063}} & \cellcolor{gray!50}{\num{1.141}} & \cellcolor{orange!50}{\num{1.059}} &  \cellcolor{gray!50}{\num{0.854}} & -/- \\\hline
\multirow{3}{*}{\cmcii} & $E_\text{g, dir}$ & \cellcolor{orange!50}{\num{0.962}} & \cellcolor{orange!50}{\num{0.950}} & \cellcolor{gray!50}{\num{1.027}} & -/- & \cellcolor{gray!50}{\num{1.146}} & \cellcolor{gray!50}{\num{1.405}} & \cellcolor{orange!50}{\num{1.013}} & \cellcolor{gray!50}{\num{0.902}} & \cellcolor{gray!50}{\num{1.841}} & \cellcolor{gray!50}{\num{1.064}} & \cellcolor{gray!50}{\num{0.852}} & \cellcolor{gray!50}{\num{0.854}} & -/- \\
& $E_\text{g, ind}$ & \cellcolor{orange!50}{-/-} & \cellcolor{orange!50}{-/-} & \cellcolor{gray!50}{\num{0.980}} & -/- & \cellcolor{gray!50}{\num{1.034}} & \cellcolor{gray!50}{\num{1.124}} & \cellcolor{orange!50}{-/-} & \cellcolor{gray!50}{\num{0.714}} & \cellcolor{gray!50}{\num{0.827}} & \cellcolor{gray!50}{\num{0.881}} & \cellcolor{gray!50}{\num{0.715}} & \cellcolor{gray!50}{\num{0.718}} & -/- \\
& $E_\text{g, opt}$ & 
\cellcolor{orange!50}{\num{1.640}} & \cellcolor{orange!50}{\num{1.370}} & \cellcolor{gray!50}{\num{1.063}} & -/- & \cellcolor{gray!50}{\num{1.169}} & \cellcolor{gray!50}{\num{1.504}} & \cellcolor{orange!50}{\num{1.420}} & \cellcolor{gray!50}{\num{0.952}} & \cellcolor{gray!50}{\num{0.853}} & \cellcolor{gray!50}{\num{1.162}} & \cellcolor{gray!50}{\num{0.854}} & \cellcolor{gray!50}{\num{0.856}} & -/- \\\hline
\multirow{3}{*}{\piic} & $E_\text{g, dir}$ & \cellcolor{gray!50}{\num{1.892}} & -/- & -/- & -/- & -/- & \cellcolor{gray!50}{\num{1.968}} & \cellcolor{gray!50}{\num{1.606}} & \cellcolor{gray!50}{\num{1.196}} & \cellcolor{gray!50}{\num{1.425}} & \cellcolor{gray!50}{\num{1.660}} & -/- & -/- & \cellcolor{gray!50}{\num{0.890}} \\
& $E_\text{g, ind}$ & \cellcolor{gray!50}{\num{1.672}} & -/- & -/- & -/- & -/- & \cellcolor{gray!50}{\num{1.717}} & \cellcolor{gray!50}{\num{1.490}} & \cellcolor{gray!50}{\num{1.076}} & \cellcolor{gray!50}{\num{1.118}} & \cellcolor{gray!50}{\num{1.474}} & -/- & -/- & \cellcolor{gray!50}{\num{0.256}} \\
& $E_\text{g, opt}$ & \cellcolor{gray!50}{\num{1.983}} & -/- & -/- & -/- & -/- & \cellcolor{gray!50}{\num{2.065}} & \cellcolor{gray!50}{\num{1.599}} & \cellcolor{gray!50}{\num{1.254}} & \cellcolor{gray!50}{\num{1.452}} & \cellcolor{gray!50}{\num{1.751}} & -/- & -/- & \cellcolor{gray!50}{\num{0.931}} \\\hline
\end{tabular}
}
\end{table}

In sum, our results indicate that the band gaps of indium-based \snbchx compounds lie in the middle and upper region of our screening range, making them interesting for both indoor and outdoor photovoltaics. However, the scarcity of indium becomes a cost factor, whereby a number of efforts are underway to ensure the long-term supply of indium through significant changes in processing and recycling. With band gaps around \SI{1}{\electronvolt} antimony and bismuth-based \snbchx materials would be interesting for outdoor photovoltaics. Considering antimony's toxicity, bismuth-based \snbchx materials might be promising contenders.

Lastly, we check which \snbchx materials are already known or have appeared in databases or datasets. Our search reveals that only three materials have been reported so far: \ch{Sn2SbS2I3},\cite{olivier1980} \ch{Sn2SbSe2I3}\cite{ibanez1984} and \ch{Sn2BiS2I3}.\cite{islam2016} For \ch{Sn2SbS2I3}, structures in all three space groups are known, but for the other two only  \cmcm and \cmcii have been reported. To the best of our knowledge we are the first to present the eight direct band gaps materials (\ch{Sn2InS2Br3}, \ch{Sn2InS2I3}, \ch{Sn2InSe2Cl3}, \ch{Sn2InSe2Br3}, \ch{Sn2InTe2Br3}, \ch{Sn2InTe2Cl3}, \ch{Sn2SbSe2Cl3} and \ch{Sn2SbTe2Cl3}) and the  12 \snbchx indirect band gaps materials (\ch{Sn2InS2Cl3}, \ch{Sn2InSe2I3}, \ch{Sn2InTe2I3}, \ch{Sn2SbS2Cl3}, \ch{Sn2SbS2Br3}, \ch{Sn2SbSe2Br3}, \ch{Sn2SbTe2Br3}, \ch{Sn2BiS2Cl3}, \ch{Sn2BiS2Br3}, \ch{Sn2BiSe2Cl3}, \ch{Sn2BiSe2Br3} and \ch{Sn2BiSe2I3}).

\subsection{Promising \protect\mixing alloys}\label{sec:results_discuss:mixing}

\begin{figure}[ht]
\includegraphics[scale=1.2125]{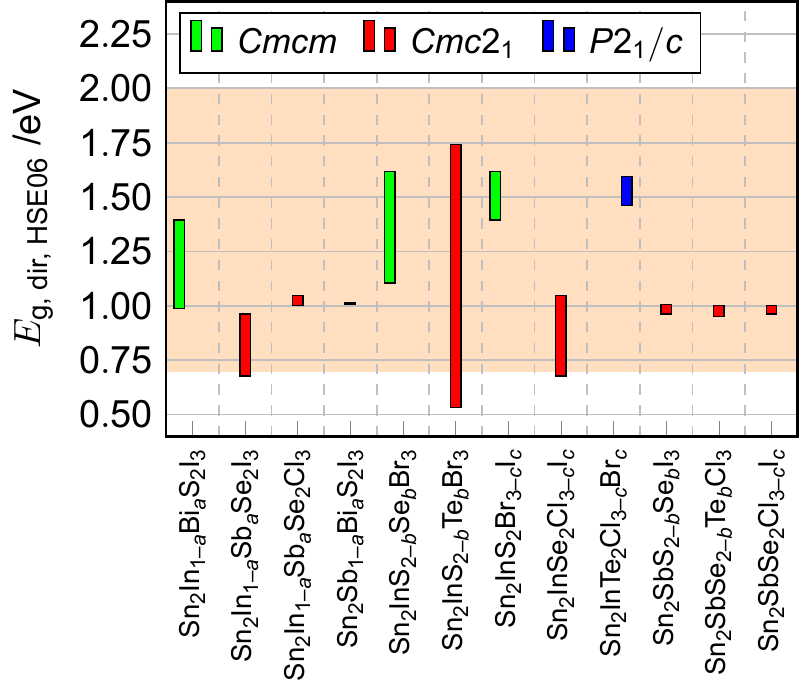}
\caption{Band gap ranges (HSE06+SOC level of theory) for the 12 \mixing binary alloys of interest for \cmcm (green), \cmcii (red) and \piic (blue) space groups.}
\label{fig:sn2bch2x3_mixing}
\end{figure}

In this section we extend the materials space to possible binary \mixing alloys with band gaps in the desired range. We consider alloys within the same space group for which I) at least one compound has a direct band gap in the target range or II) one compound has a direct band gap below \SI{0.7}{\electronvolt} and the another above \SI{2}{\electronvolt}. We identified 12 \mixing suitable alloys (see \cref{fig:sn2bch2x3_mixing}).  We expect direct band gaps in at least a part of the compositions. With such alloys, the band gap could be tuned  for specific optoelectronic applications. To the best of our knowledge, none of these binary \mixing alloy compounds have been studied in the literature yet.

We find that the \cmcii space group is most prevalent followed by \cmcm. Only a single binary alloy has the \piic space group because of the phase's preference for indirect gaps. Materials in the \cmcii space group tend towards lower band gaps close to $\sim $\SI{1}{\electronvolt}, because the parent compounds have small gaps. One exception is \ch{Sn2InS_{2-{$b$}}Te_{$b$}Br3} which combines \ch{Sn2InTe2Br3} ($E_\text{g} = \SI{0.531}{\electronvolt}$) and \ch{Sn2InTe2Br3} ($E_\text{g} = \SI{1.742}{\electronvolt}$), resulting in a broad range of \SI{1.211}{\electronvolt}. For the three \cmcm alloys, we find \ch{Sn2In_{1-{$a$}}Bi_{$a$}S2I3} at the lower end, \ch{Sn2InS_{2-{$b$}}Se_{$b$}Br3} and \ch{Sn2InS2Br_{3-{$c$}}Cl_{$c$}} in the middle of the range. If we drop our restriction of mixing compounds with a direct gap, we obtain an additional 69 \mixing alloys reported in Figure\,4 of the SI.

Our results indicate that \mixing alloys with indium and iodine offer the most flexibility (i.e. widest tuning ranges). However, considering the scarcity of indium and the relatively high toxicity of antimony, alloys with bismuth would be favoured. Since we could not identify a suitable bismuth alloy, \ch{Sn2In_{1-{$a$}}Bi_{$a$}S2I3} might offer the best, environmentally friendly compromise.

\section{Conclusion}\label{sec:conclusion}

In the pursuit of low-toxicity photovoltaic materials, we used DFT to explore the \snbchx material space. We identified 12 absorbers that fulfill our requirements by being a lead-free, thermodynamically stable and with a direct band gap in the screening range of \SIrange{0.7}{2.0}{\electronvolt}. Three of these \snbchx were known previously, while the remaining nine materials, i.e., \ch{Sn2InS2Br3}, \ch{Sn2InS2I3}, \ch{Sn2InSe2Cl3}, \ch{Sn2InSe2Br3}, \ch{Sn2InTe2Br3}, \ch{Sn2InTe2Cl3}, \ch{Sn2SbSe2Cl3}, \ch{Sn2SbTe2Cl3} and \ch{Sn2BiTe2Cl3}, are reported here for the first time. The majority of these compounds contain indium, including some of the most promising candidates, like \ch{Sn2InSe2Br3} or \ch{Sn2InS2Br3}. An alternative to indium- and antimony-based \snbchx materials are bismuth-based ones, such as e.g., \ch{Sn2BiTe2Cl3}, although the band gap is smaller than its indium-based counterparts. In addition, we identified 15 previously unknown \snbchx compounds with indirect band gaps, which have a lowest direct band gap within the screening range. 

We investigated the potential to tune the band gap by exploring \mixing alloys. In total, 12 \mixing alloys were identified by alloying within the same phase and considering only structures with direct gaps. For example, the fundamental gap of \ch{SnIn_{1-{$a$}}Bi_{$a$}S2Br3} can be tuned between \SIrange{0.988}{1.395}{\electronvolt}. Furthermore, we identified additional 69 \mixing alloys by dropping our requirement to use only direct band gap \snbchx end structures, which broadens significantly the \mixing spectra. 

Our findings highlight the relevance of materials design and theoretical explorations, and the identified \snbchx materials and \mixing alloys encourage the synthesis and the use of highly stable materials in various optoelectronic devices.

\newpage

\begin{acknowledgement}
The authors thank Milica Todorovi\'c and Armi Tiihonen for fruitful discussions. This study was supported by the Academy of Finland through Project No. 334532 and the National Natural Science Foundation of China (Grant No. 62281330043). PH, JL and PR further acknowledge CSC-IT Center for Science, Finland, the Aalto Science-IT project, Xi'an Jiaotong University's HPC platform and the Hefei Advanced Computing Center of China for generous computational resources. PV acknowledges the financial support of Academy of Finland, Decision No. 347772. This work is part of the Academy of Finland Flagship Programme, Photonics Research and Innovation (PREIN), Decision No. 320165. GKG acknowledges Tampere Institute for Advanced Study for postdoctoral funding.
\end{acknowledgement}

\begin{suppinfo}
The Supporting Information is available free of charge at \ldots
\begin{itemize}
  \item band structure for \ch{Sn2InS2Br3} (\cmcm)
  \item lattice parameters, formation energies, fundamental (diract and indirect) and optical gaps for all \snbchx materials
  \item formation energies of indirect \snbchx materials
  \item band gaps for all \snbchx materials and \mixing alloys
\end{itemize}
\end{suppinfo}

\section*{Notes}
The authors declare no competing financial interest.

\bibliography{lit_a2bch2x3}

\end{document}


\newpage
\pdfbookmark[section]{Band structure of \protect\ch{Sn2InS2Br3}}{s1}
\section{S1: Band structure of \protect\ch{Sn2InS2Br3}}

\begin{figure}[!]
\centering
\includegraphics[scale=.325]{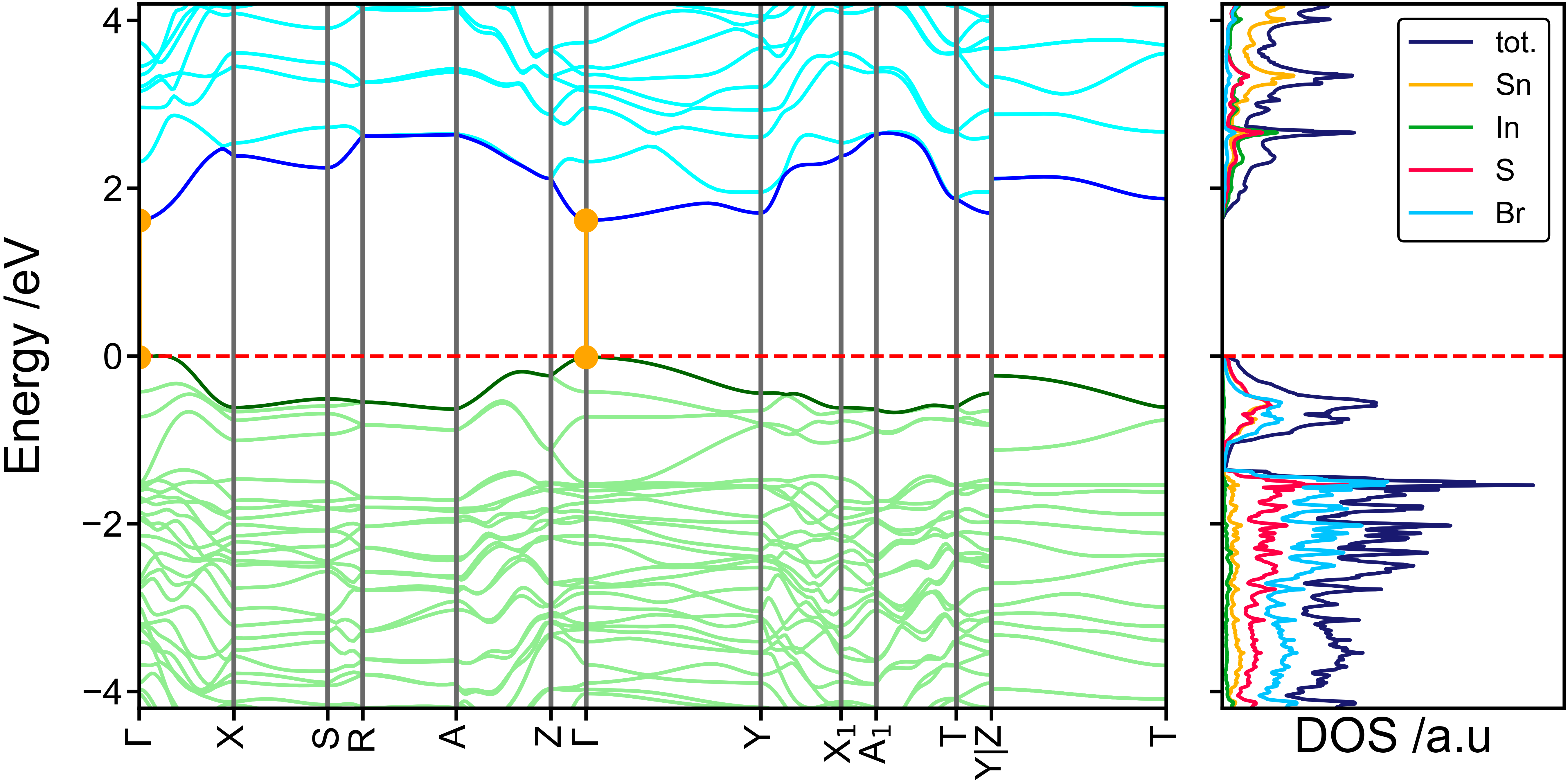}
\caption{Bandstructure (left) and normalized density of states (right) for \ch{Sn2InS2Br3} for the $Cmcm$ space group. Conduction bands are shown in green and valence bands in blue. The total and partial denstiy of state per element are provided.}
\end{figure}

\newpage
\pdfbookmark[section]{Lattice parameters, formation energies, fundamental and optical band gaps for \protect\snbchx materials}{s2}
\section{S2: Lattice parameters, formation energies, fundamental and optical band gaps for \protect\snbchx materials}

\vspace*{-0.05cm}

\begin{table}[h!]
\centering
\caption{Summary of the calculated lattice parameters (\textit{a}, \textit{b}, \textit{c}, \textit{a/c}-ratio), as well as the formation energy $E_\text{form}$ (related to the $i$-th elements) using PBEsol level of theory and the fundamental lowest direct band gap $E_\text{g. dir.}$, and for indirect \snbchx materials also the lowest indirect band gap ($E_\text{g, ind.}$), as well as the optical band gap using HSE06+SOC level of theory. -/- marks that either the material has no indirect/no band gap at all or that no experimental values are available. \snbchx materials highlighted orange (direct) or gray (indirect) if their fundamental band gap is within the range of \SIrange{0.7}{2.0}{\electronvolt}.}
\label{si:tab:lpara_forme_gaps_sn2bch2x3:direct}
\scalebox{.75}{
\begin{tabular}{c|c|cccc|c|cc|c}
& & \multicolumn{4}{c}{\textbf{lat. parameters}} & & & \\
\textbf{System} & \textbf{Group} & \textit{a} /\si{\angstrom} & \textit{b} /\si{\angstrom} & \textit{c} /\si{\angstrom} & \textit{c/a} & $E_\text{form}$ /\si{\electronvolt\per\at} & $E_\text{g, dir.}$ & $E_\text{g, ind.}$ /\si{\electronvolt} & $E_\text{g, opt.}$ /\si{\electronvolt} \\\hline
\rowcolor{gray!50}
\cellcolor{white}{} & \cmcm & \num{4.23} & \num{14.12} & \num{14.3} & \num{3.38} & \num{-726} & \num{1.305} & \num{1.251} & \num{1.49} \\
\rowcolor{gray!50}
\cellcolor{white}{\ch{Sn2InS2Cl3}} & \cmcii & \num{3.94} & \num{14.22} & \num{15.0} & \num{3.81} & \num{-772} & \num{1.834} & \num{1.692} & \num{1.89} \\
\cellcolor{white}{} & \piic & \num{6.27} & \num{16.19} & \num{7.88} & \num{1.26} & \num{-779} & \num{2.433} & \num{2.037} & \num{2.440} \\\hline
\rowcolor{orange!50}
\cellcolor{white}{} & \cmcm & \num{4.22} & \num{14.80} & \num{14.50} & \num{3.44} & \num{-650} & \num{1.617} & -/- & \num{1.977} \\
\rowcolor{orange!50}
\cellcolor{white}{\ch{Sn2InS2Br3}} & \cmcii & \num{4.11} & \num{15.16} & \num{15.53} & \num{3.77} & \num{-693} & \num{1.742} & -/- & \num{1.783} \\
\rowcolor{gray!50}
\cellcolor{white}{} & \piic & \num{6.36} & \num{16.55} & \num{8.07} & \num{1.27} & \num{-690} & \num{1.985} & \num{1.616} & \num{1.996} \\\hline
\rowcolor{orange!50}
\cellcolor{white}{} & \cmcm & \num{4.37} & \num{15.68} & \num{14.83} & \num{3.40} & \num{-520} & \num{1.395} & -/- & \num{1.934} \\
 \rowcolor{gray!50}
\cellcolor{white}{\ch{Sn2InS2I3}} & \cmcii & \num{4.06} & \num{14.13} & \num{19.25} & \num{4.75} & \num{-519} & \num{1.120} & \num{0.598} & \num{1.187} \\
\cellcolor{white}{} & \piic & \num{7.66} & \num{16.56} & \num{8.41} & \num{1.1} & \num{-565} & \num{2.278} & \num{2.102} & \num{2.313} \\\hline
\cellcolor{white}{} & \cmcm & \num{4.13} & \num{14.94} & \num{14.9} & \num{3.60} & \num{-718} & \num{0.661} & \num{0.597} & \num{0.675} \\
 \rowcolor{orange!50}
\cellcolor{white}{\ch{Sn2InSe2Cl3}} & \cmcii & \num{4.08} & \num{14.09} & \num{16.02} & \num{3.93} & \num{-767} & \num{1.046} & -/- & \num{1.684} \\
\cellcolor{white}{} & \piic & \num{6.35} & \num{16.79} & \num{8.09} & \num{1.27} & \num{-765} & \num{2.076} & \num{1.925} & \num{2.088} \\\hline
\rowcolor{orange!50}
\cellcolor{white}{} & \cmcm & \num{4.42} & \num{14.54} & \num{15.01} & \num{3.40} & \num{-651} & \num{1.105} & -/- & \num{1.477} \\
\rowcolor{gray!50}
\cellcolor{white}{\ch{Sn2InSe2Br3}} & \cmcii & \num{3.98} & \num{15.22} & \num{15.58} & \num{3.91} & \num{-699} & \num{1.373} & \num{1.338} & \num{1.396} \\
\rowcolor{gray!50}
\cellcolor{white}{} & \piic & \num{6.41} & \num{17.09} & \num{8.23} & \num{1.28} & \num{-688} & \num{1.843} & \num{1.488} & \num{1.855} \\\hline
\rowcolor{gray!50}
\cellcolor{white}{} & \cmcm & \num{4.39} & \num{15.73} & \num{15.31} & \num{3.49} & \num{-534} & \num{1.369} & \num{1.299} & \num{1.397} \\
\cellcolor{white}{\ch{Sn2InSe2I3}} & \cmcii & \num{4.32} & \num{15.93} & \num{16.32} & \num{3.78} & \num{-577} & \num{0.677} & -/- & \num{1.282} \\
\rowcolor{gray!50}
\cellcolor{white}{} & \piic & \num{7.41} & \num{17.62} & \num{8.5} & \num{1.15} & \num{-567} & \num{1.947} & \num{1.785} & \num{1.960} \\\hline
\cellcolor{white}{} & \cmcm & \num{4.19} & \num{12.25} & \num{18.4} & \num{4.39} & \num{-542} & \num{0.558} & \num{0.455} & \num{0.651} \\
\rowcolor{gray!50}
\cellcolor{white}{\ch{Sn2InTe2Cl3}} & \cmcii & \num{4.16} & \num{13.24} & \num{17.21} & \num{4.13} & \num{-639} & \num{0.915} & \num{0.804} & \num{1.011} \\
\rowcolor{orange!50}
\cellcolor{white}{} & \piic & \num{6.53} & \num{17.82} & \num{8.58} & \num{1.32} & \num{-610} & \num{1.594} & -/- & \num{1.713} \\\hline
\cellcolor{white}{} & \cmcm & \num{4.28} & \num{12.34} & \num{18.41} & \num{4.30} & \num{-488} & -/- & -/- & -/- \\
\cellcolor{white}{\ch{Sn2InTe2Br3}} & \cmcii & \num{4.31} & \num{14.70} & \num{16.90} & \num{3.92} & \num{-559} & \num{0.531} & -/- & \num{0.786} \\
\rowcolor{orange!50}
\cellcolor{white}{} & \piic & \num{6.48} & \num{18.04} & \num{8.60} & \num{1.33} & \num{-543} & \num{1.461} & -/- & \num{1.592} \\\hline
\rowcolor{gray!50}
\cellcolor{white}{} & \cmcm & \num{4.56} & \num{15.6} & \num{16.26} & \num{3.57} & \num{-425} & \num{1.089} & \num{0.994} & \num{1.126} \\
\rowcolor{gray!50}
\cellcolor{white}{\ch{Sn2InTe2I3}} & \cmcii & \num{4.3} & \num{15.34} & \num{16.92} & \num{3.94} & \num{-460} & \num{0.867} & \num{0.802} & \num{0.964} \\
\rowcolor{gray!50}
\cellcolor{white}{} & \piic & \num{6.58} & \num{18.87} & \num{8.74} & \num{1.33} & \num{-438} & \num{1.109} & \num{1.083} & \num{1.201} \\\hline

\rowcolor{gray!50}
\cellcolor{white}{} & \cmcm & \num{4.08} & \num{12.63} & \num{15.88} & \num{3.89} & \num{-649} & \num{1.679} & \num{1.611} & \num{1.773} \\
\rowcolor{gray!50}
\cellcolor{white}{\ch{Sn2SbS2Cl3}} & \cmcii & \num{4.01} & \num{13.81} & \num{16.0} & \num{3.99} & \num{-655} & \num{1.242} & \num{1.056} & \num{1.288} \\
\cellcolor{white}{} & \piic & \num{6.96} & \num{16.05} & \num{8.08} & \num{1.16} & \num{-702} & \num{2.451} & \num{2.170} & \num{2.477} \\\hline
\rowcolor{gray!50}
\cellcolor{white}{} & \cmcm & \num{4.1} & \num{13.47} & \num{15.4} & \num{3.76} & \num{-574} & \num{1.474} & \num{0.883} & \num{1.527} \\
\rowcolor{gray!50}
\cellcolor{white}{\ch{Sn2SbS2Br3}} & \cmcii & \num{4.1} & \num{13.92} & \num{16.12} & \num{3.93} & \num{-592} & \num{1.248} & \num{1.055} & \num{1.344} \\
\cellcolor{white}{} & \piic & \num{7.01} & \num{16.06} & \num{8.19} & \num{1.17} & \num{-646} & \num{2.194} & \num{1.989} & \num{2.294} \\\hline
\rowcolor{gray!50}
\cellcolor{white}{} & \cmcm & \num{4.24} & \num{13.84} & \num{15.76} & \num{3.72} & \num{-486} & \num{0.997} & \num{0.947} & \num{1.547} \\
\multirow{2}{*}{\ch{Sn2SbS2I3}} & \cellcolor{orange!50}{\cmcii} & \cellcolor{orange!50}{\num{4.26}} & \cellcolor{orange!50}{\num{14.22}} & \cellcolor{orange!50}{\num{16.24}} & \cellcolor{orange!50}{\num{3.81}} & \cellcolor{orange!50}{\num{-497}} & \cellcolor{orange!50}{\num{1.007}} & \cellcolor{orange!50}{-/-} & \cellcolor{orange!50}{\num{1.506}} \\
 & \cellcolor{gray!50}{\piic} & \cellcolor{gray!50}{\num{7.25}} & \cellcolor{gray!50}{\num{16.24}} & \cellcolor{gray!50}{\num{8.47}} & \cellcolor{gray!50}{\num{1.17}} & \cellcolor{gray!50}{\num{-555}} & \cellcolor{gray!50}{\num{1.762}} & \cellcolor{gray!50}{\num{1.686}} & \cellcolor{gray!50}{\num{1.771}} \\
 & Exp.\textsuperscript{\hyperref[fn:tab_a_bottom]{\textit{a}}\protect\phantomsection\label{fn:tab_a_top}} & \num{4.25} & \num{13.99} & \num{16.38} & \num{3.85} & -/- & -/- & -/- & \num{1.41} \\\hline
\rowcolor{orange!50}
\cellcolor{white}{} & \cmcm & \num{4.16} & \num{12.48} & \num{16.92} & \num{4.07} & \num{-646} & \num{1.411} & -/- & \num{1.816}\\
\rowcolor{orange!50}
\cellcolor{white}{\ch{Sn2SbSe2Cl3}} & \cmcii & \num{4.07} & \num{14.01} & \num{16.45} & \num{4.04} & \num{-651} & \num{1.002} & -/- & \num{1.261} \\
\cellcolor{white}{} & \piic & \num{7.06} & \num{17.98} & \num{8.06} & \num{1.14} & \num{-692} & \num{2.181} & \num{1.839} & \num{2.202} \\\hline
\end{tabular}}
\end{table}

\newpage

\begin{table}[h!]
\begin{minipage}{\textwidth}
\centering
\scalebox{.75}{
\begin{tabular}{c|c|cccc|c|cc|c}
& & \multicolumn{4}{c}{\textbf{lat. parameters}} & & & \\
\textbf{System} & \textbf{Group} & \textit{a} /\si{\angstrom} & \textit{b} /\si{\angstrom} & \textit{c} /\si{\angstrom} & \textit{c/a} & $E_\text{form}$ /\si{\electronvolt\per\at} & $E_\text{g, dir.}$ & $E_\text{g, ind.}$ /\si{\electronvolt} & $E_\text{g, opt.}$ /\si{\electronvolt} \\\hline
\rowcolor{gray!50}
\cellcolor{white}{} & \cmcm & \num{4.13} & \num{12.94} & \num{17.75} & \num{4.3} & \num{-572} & \num{1.412} & \num{1.330} & \num{1.504} \\
\rowcolor{gray!50}
\cellcolor{white}{\ch{Sn2SbSe2Br3}} & \cmcii & \num{4.16} & \num{14.22} & \num{16.73} & \num{4.03} & \num{-586} & \num{0.941} & \num{0.891} & \num{1.044} \\
\rowcolor{gray!50}
\cellcolor{white}{} & \piic & \num{7.1} & \num{17.16} & \num{8.29} & \num{1.17} & \num{-631} & \num{2.028} & \num{1.642} & \num{2.100} \\\hline
\rowcolor{gray!50}
\cellcolor{white}{} & \cmcm & \num{4.28} & \num{13.95} & \num{16.44} & \num{3.84} & \num{-473} & \num{1.057} & \num{0.800} & \num{1.056} \\
\multirow{2}{*}{\ch{Sn2SbSe2I3}} & \cellcolor{orange!50}{\cmcii} & \cellcolor{orange!50}{\num{4.31}} & \cellcolor{orange!50}{\num{14.50}} & \cellcolor{orange!50}{\num{16.91}} & \cellcolor{orange!50}{\num{3.93}} & \cellcolor{orange!50}{\num{-493}} & \cellcolor{orange!50}{\num{0.962}} & \cellcolor{orange!50}{-/-} & \cellcolor{orange!50}{\num{1.640}} \\
 & \cellcolor{gray!50}{\piic} & \cellcolor{gray!50}{\num{7.28}} & \cellcolor{gray!50}{\num{17.08}} & \cellcolor{gray!50}{\num{8.57}} & \cellcolor{gray!50}{\num{1.18}} & \cellcolor{gray!50}{\num{-547}} & \cellcolor{gray!50}{\num{1.892}} & \cellcolor{gray!50}{\num{1.672}} & \cellcolor{gray!50}{\num{1.983}} \\
\cellcolor{white}{} & Exp.\textsuperscript{\hyperref[fn:tab_b_bottom]{\textit{b}}\protect\phantomsection\label{fn:tab_b_top}} & \num{4.30} & \num{14.09} & \num{17.22} & \num{4.01} & -/- & -/- & -/- & -/- \\\hline
\rowcolor{gray!50}
\cellcolor{white}{} & \cmcm & \num{4.34} & \num{12.40} & \num{17.68} & \num{4.07} & \num{-530} & \num{0.943} & \num{0.874} & \num{0.997} \\
\rowcolor{orange!50}
\cellcolor{white}{\ch{Sn2SbTe2Cl3}} & \cmcii & \num{4.34} & \num{12.39} & \num{17.70} & \num{4.08} & \num{-523} & \num{0.950} & -/- & \num{1.370} \\
\cellcolor{white}{} & \piic & \num{6.45} & \num{18.1} & \num{8.36} & \num{1.30} & \num{-588} & \num{0.559} & \num{0.480} & \num{0.653} \\\hline
\cellcolor{white}{} & \cmcm & \num{4.24} & \num{12.51} & \num{19.69} & \num{4.64} & \num{-456} & \num{0.699} & \num{0.490} & \num{0.763} \\
\rowcolor{gray!50}
\cellcolor{white}{\ch{Sn2SbTe2Br3}} & \cmcii & \num{4.27} & \num{14.66} & \num{17.39} & \num{4.07} & \num{-456} & \num{1.027} & \num{0.980} & \num{1.063} \\
\cellcolor{white}{} & \piic & \num{6.59} & \num{18.5} & \num{8.42} & \num{1.28} & \num{-518} & \num{0.545} & \num{0.245} & \num{0.567} \\\hline
\rowcolor{gray!50}
\cellcolor{white}{} & \cmcm & \num{4.34} & \num{13.57} & \num{19.47} & \num{4.48} & \num{-353} & \num{0.786} & \num{0.465} & \num{0.795} \\
\cellcolor{white}{\ch{Sn2SbTe2I3}} & \cmcii & \num{4.28} & \num{14.31} & \num{18.73} & \num{4.37} & \num{-358} & \num{0.376} & \num{0.186} & \num{0.433} \\
\cellcolor{white}{} & \piic & \num{7.02} & \num{18.82} & \num{8.67} & \num{1.24} & \num{-412} & \num{0.852} & \num{0.021} & \num{1.255} \\\hline

\rowcolor{gray!50}
\cellcolor{white}{} & \cmcm & \num{3.99} & \num{13.91} & \num{16.02} & \num{4.02} & \num{-714} & \num{1.535} & \num{0.986} & \num{1.548} \\
\rowcolor{gray!50}
\cellcolor{white}{\ch{Sn2BiS2Cl3}} & \cmcii & \num{4.07} & \num{13.72} & \num{15.98} & \num{3.93} & \num{-716} & \num{1.146} & \num{1.034} & \num{1.169} \\
\cellcolor{white}{} & \piic & \num{6.97} & \num{15.93} & \num{8.1} & \num{1.16} & \num{-732} & \num{2.187} & \num{1.84} & \num{2.191} \\\hline
 
\rowcolor{gray!50}
\cellcolor{white}{} & \cmcm & \num{4.11} & \num{13.42} & \num{15.68} & \num{3.81} & \num{-650} & \num{1.343} & \num{0.956} & \num{1.434} \\
\rowcolor{gray!50}
\cellcolor{white}{\ch{Sn2BiS2Br3}} & \cmcii & \num{4.11} & \num{13.42} & \num{15.68} & \num{3.82} & \num{-651} & \num{1.405} & \num{1.124} & \num{1.504} \\
\rowcolor{gray!50}
\cellcolor{white}{} & \piic & \num{7.03} & \num{16.04} & \num{8.23} & \num{1.17} & \num{-675} & \num{1.968} & \num{1.717} & \num{2.065} \\\hline
\rowcolor{orange!50}
\cellcolor{white}{} & \cmcm & \num{4.26} & \num{13.92} & \num{15.93} & \num{3.74} & \num{-552} & \num{0.988} & -/- & \num{1.390} \\
\multirow{2}{*}{\ch{Sn2BiS2I3}} & \cellcolor{orange!50}{\cmcii} & \cellcolor{orange!50}{\num{4.26}} & \cellcolor{orange!50}{\num{13.92}} & \cellcolor{orange!50}{\num{15.94}} & \cellcolor{orange!50}{\num{3.74}} & \cellcolor{orange!50}{\num{-553}} & \cellcolor{orange!50}{\num{1.013}} & \cellcolor{orange!50}{-/-} & \cellcolor{orange!50}{\num{1.420}} \\
 & \cellcolor{gray!50}{\piic} & \cellcolor{gray!50}{\num{7.3}} & \cellcolor{gray!50}{\num{16.23}} & \cellcolor{gray!50}{\num{8.52}} & \cellcolor{gray!50}{\num{1.17}} & \cellcolor{gray!50}{\num{-583}} & \cellcolor{gray!50}{\num{1.606}} & \cellcolor{gray!50}{\num{1.490}} & \cellcolor{gray!50}{\num{1.599}} \\
 & Exp.\textsuperscript{\hyperref[fn:tab_c_bottom]{\textit{c}}\protect\phantomsection\label{fn:tab_c_top}} & \num{4.29} & \num{14.12} & \num{16.41} & \num{3.83} & -/- & -/- & -/- & \num{1.22} \\\hline
\rowcolor{gray!50}
\cellcolor{white}{} & \cmcm & \num{4.03} & \num{12.5} & \num{17.07} & \num{4.23} & \num{-713} & \num{1.104} & \num{1.096} & \num{1.124} \\
\rowcolor{gray!50}
\cellcolor{white}{\ch{Sn2BiSe2Cl3}} & \cmcii & \num{4.13} & \num{13.98} & \num{16.46} & \num{3.99} & \num{-708} & \num{0.902} & \num{0.714} & \num{0.952} \\
\rowcolor{gray!50}
\cellcolor{white}{} & \piic & \num{6.41} & \num{17.23} & \num{8.22} & \num{1.28} & \num{-731} & \num{1.196} & \num{1.076} & \num{1.254} \\\hline
\rowcolor{gray!50}
\cellcolor{white}{} & \cmcm & \num{4.13} & \num{13.02} & \num{17.29} & \num{4.19} & \num{-635} & \num{0.965} & \num{0.666} & \num{1.063} \\
\rowcolor{gray!50}
\cellcolor{white}{\ch{Sn2BiSe2Br3}} & \cmcii & \num{4.21} & \num{14.14} & \num{16.74} & \num{3.98} & \num{-642} & \num{0.841} & \num{0.827} & \num{0.853} \\
\rowcolor{gray!50}
\cellcolor{white}{} & \piic & \num{6.9} & \num{17.19} & \num{8.33} & \num{1.21} & \num{-662} & \num{1.425} & \num{1.118} & \num{1.452} \\\hline
\rowcolor{gray!50}
\cellcolor{white}{} & \cmcm & \num{4.31} & \num{13.99} & \num{16.62} & \num{3.86} & \num{-538} & \num{1.048} & \num{0.769} & \num{1.141} \\
\rowcolor{gray!50}
\cellcolor{white}{\ch{Sn2BiSe2I3}} & \cmcii & \num{4.35} & \num{14.47} & \num{16.87} & \num{3.87} & \num{-538} & \num{1.064} & \num{0.881} & \num{1.162} \\
\rowcolor{gray!50}
\cellcolor{white}{} & \piic & \num{7.33} & \num{17.08} & \num{8.6} & \num{1.17} & \num{-575} & \num{1.66} & \num{1.474} & \num{1.751} \\\hline
\rowcolor{orange!50}
\cellcolor{white}{} & \cmcm & \num{4.15} & \num{12.37} & \num{17.83} & \num{4.30} & \num{-580} & \num{0.738} & -/- & \num{1.079} \\
\rowcolor{gray!50}
\cellcolor{white}{\ch{Sn2BiTe2Cl3}} & \cmcii & \num{4.20} & \num{12.37} & \num{17.84} & \num{4.25} & \num{-579} & \num{0.764} & \num{0.697} & \num{0.863} \\
\cellcolor{white}{} & \piic & \num{6.53} & \num{18.02} & \num{8.48} & \num{1.30} & \num{-612} & \num{0.498} & \num{0.476} & \num{0.579} \\\hline
\rowcolor{gray!50}
\cellcolor{white}{} & \cmcm & \num{4.24} & \num{12.63} & \num{18.26} & \num{4.31} & \num{-510} & \num{0.852} & \num{0.715} & \num{0.854} \\
\rowcolor{gray!50}
\cellcolor{white}{\ch{Sn2BiTe2Br3}} & \cmcii & \num{4.24} & \num{12.64} & \num{18.27} & \num{4.31} & \num{-510} & \num{0.854} & \num{0.718} & \num{0.856} \\
\cellcolor{white}{} & \piic & \num{6.68} & \num{18.39} & \num{8.53} & \num{1.28} & \num{-542} & \num{0.597} & \num{0.440} & \num{0.692} \\\hline
\cellcolor{white}{} & \cmcm & \num{4.38} & \num{13.30} & \num{18.73} & \num{4.28} & \num{-399} & -/- & -/- & -/- \\
\cellcolor{white}{\ch{Sn2BiTe2I3}} & \cmcii & \num{4.48} & \num{14.90} & \num{17.96} & \num{4.05} & \num{-407} & \num{0.610} & \num{0.582} & \num{0.692} \\
\rowcolor{gray!50}
\cellcolor{white}{} & \piic & \num{7.12} & \num{18.67} & \num{8.75} & \num{1.23} & \num{-439} & \num{0.890} & \num{0.256} & \num{0.931} \\\hline
\end{tabular}}
\vspace*{.17cm}
\footnoterule\footnotesize
\vspace*{-.5cm}
\begin{flushleft}
\textsuperscript{\hyperref[fn:tab_a_top]{\textit{a}}}\hspace*{0.0545cm} Lat. para.  are taken from RT ($T = \SI{293}{\kelvin}$) XRD meas. using Mo$K\alpha$ radiation\cite{ibanez1984} and band gap are taken from UV/vis absorption meas.\cite{nie2020} \label{fn:tab_a_bottom}\\[.25cm]
\textsuperscript{\hyperref[fn:tab_b_top]{\textit{b}}}\hspace*{0.0545cm} Lat. para. are taken from low-temp. ($T = \SI{173}{\kelvin}$) XRD meas. using Mo$K\alpha$ radiation. \cite{ibanez1984} \label{fn:tab_b_bottom}\\[.25cm]
\textsuperscript{\hyperref[fn:tab_c_top]{\textit{c}}}\hspace*{0.0545cm} Lat. para.  are taken from RT ($T = \SI{293}{\kelvin}$) XRD meas. using Mo$K\alpha$ radiation and band gap are taken from UV/vis absorption meas.\cite{islam2016} \label{fn:tab_c_bottom}
\end{flushleft}
\end{minipage}
\end{table}

\clearpage

\pdfbookmark[section]{Formation energies for indirect \protect\snbchx \mbox{materials}}{s3}
\section{S3: Formation energies for indirect \protect\snbchx \mbox{materials}}

\begin{figure}[h!]
\includegraphics[scale=1.25]{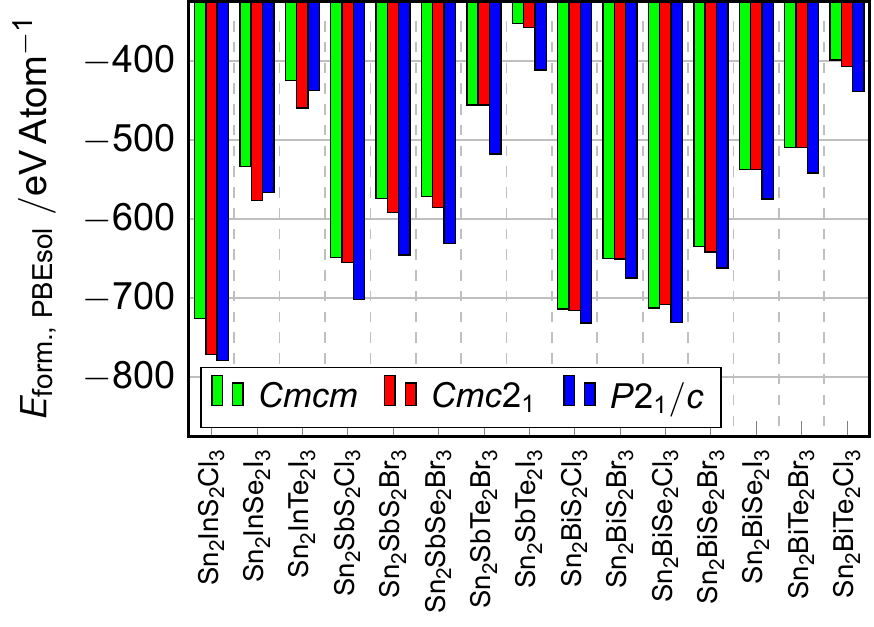}
\caption{Formation energy (calculated with PBEsol) of the 15 indirect \snbchx compounds of interest for \cmcm (green), \cmcii (red) and \piic (blue) space groups.}
\label{si:fig:indirect_sn2bch2x3_formationenergy}
\end{figure}

\clearpage

\pdfbookmark[section]{Band gaps for all identified direct and indirect \protect\snbchx materials}{s4}
\section{S4: Band gaps for all identified direct and indirect \protect\snbchx materials}

\begin{figure}[h!]
\includegraphics[scale=1.25]{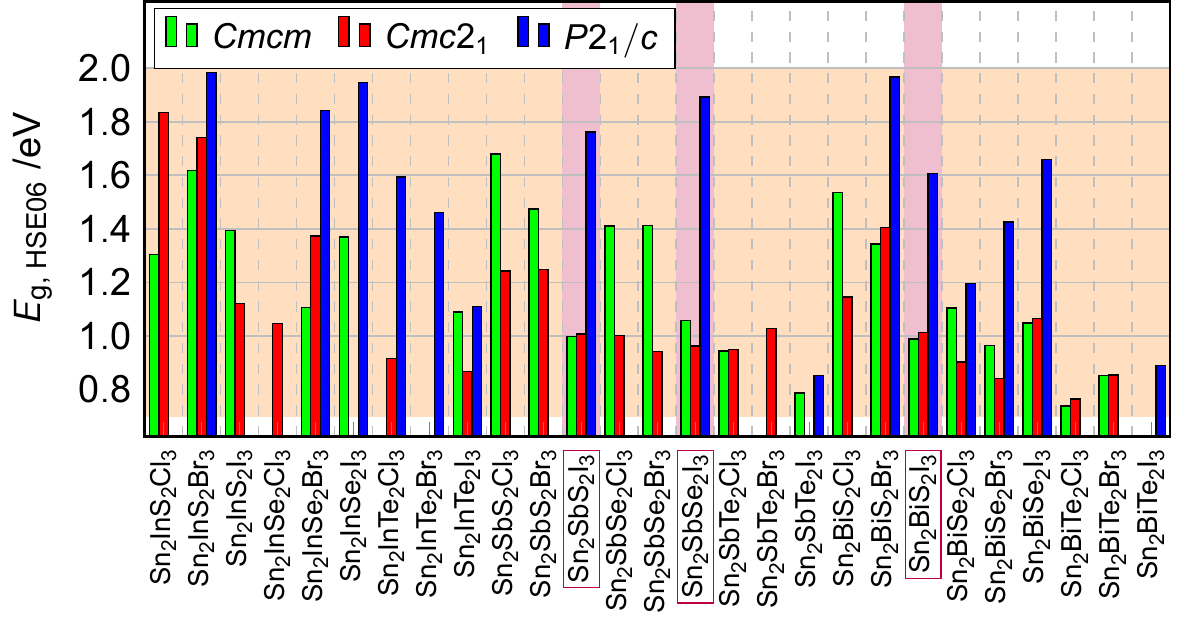}
\caption{Direct band gaps (HSE06+SOC level of theory) for 12 \ch{Sn2SbS2I3} materials of interest and energetically lowest direct band gaps (HSE06+SOC level of theory) for 15 indirect \ch{Sn2SbS2I3} materials of interest for \cmcm (green), \cmcii (red) and \piic (blue) space groups. Three structures - \ch{Sn2SbS2I3},\cite{olivier1980} \ch{Sn2SbSe2I3},\cite{ibanez1984} and \ch{Sn2BiS2I3}\cite{islam2016} - that are already known from the literature, are highlighted in purple.}
\label{si:fig:sn2bch2x3_fundbandgap_direct_indirect}
\end{figure}

\clearpage

\pdfbookmark[section]{\protect\mixing alloy compounds based on direct and indirect \protect\snbchx materials}{s5}
\section{S5: \protect\mixing alloy compounds based on direct and indirect \protect\snbchx materials}

\begin{figure}[h!]
\includegraphics[scale=0.80]{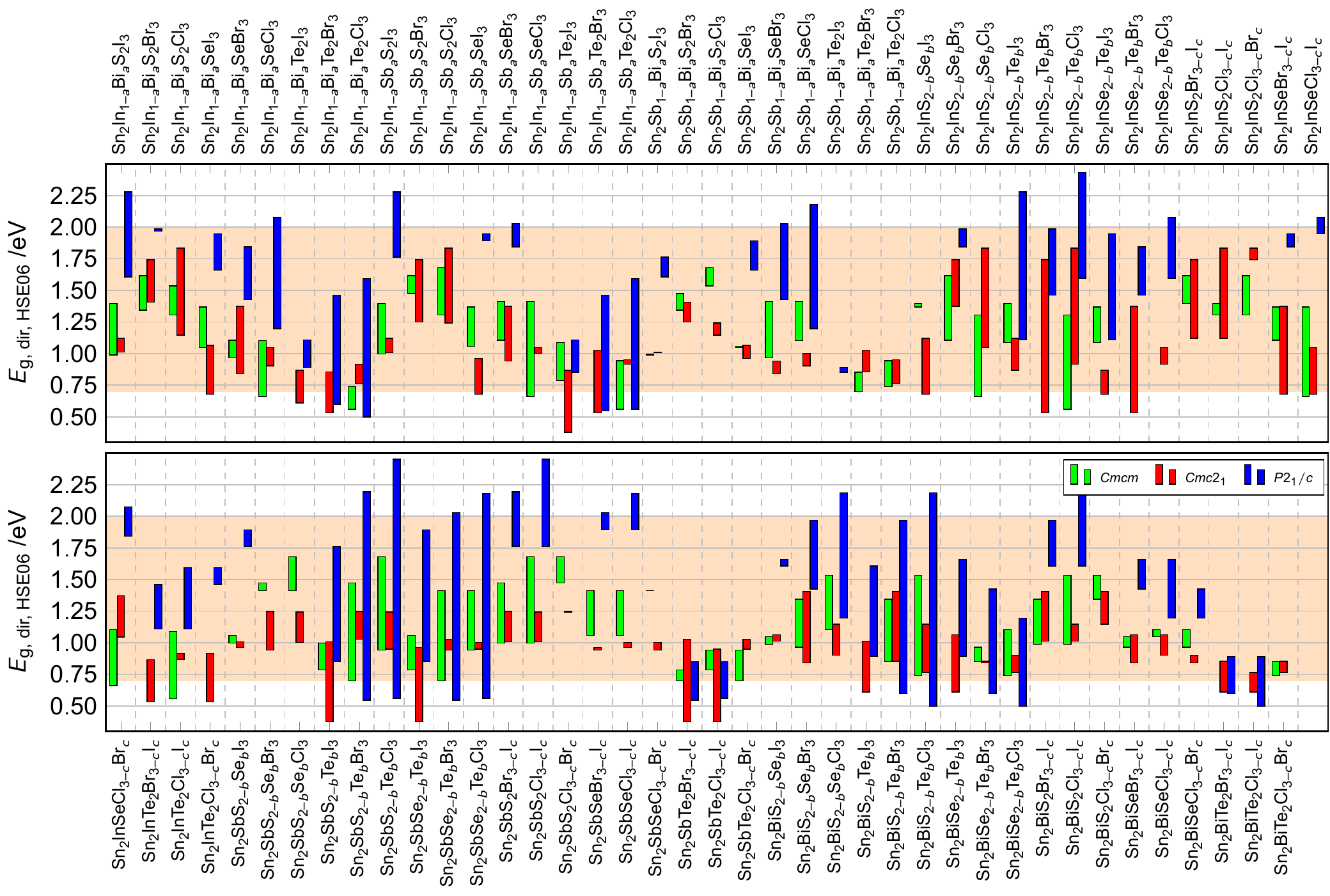}
\caption{Band gap ranges (HSE06+SOC level of theory) for all 81 \mixing binary alloy compounds of interest based on direct an indirect \snbchx compounds for \cmcm (green), \cmcii (red) and \piic (blue) space groups.}
\label{si:fig:sn2bch2x3_mixing_all}
\end{figure}

Similar to the described approach of \mixing alloys with only \snbchx materials with direct gaps, we now consider also \snbchx materials with an indirect gaps. In this way, we obtain a considerably larger \mixing material space. The alloys we consider are composed either of two direct \snbchx materials, or of one direct and one indirect material, as well as of two indirect materials. For the indirect \snbchx compounds, we use the energetically lowest direct band gap, analogous to \cref{si:fig:sn2bch2x3_fundbandgap_direct_indirect} in the SI. By this we can identify a total of 81 \mixing alloys [see \cref{si:fig:sn2bch2x3_mixing_all} in the SI], 69 of which are composed of at least one indirect \snbchx material.

The \piic space group is no longer underrepresented and a homogenous distribution of the three space groups is obtained. Similar to their direct counterparts, the alloys within the \cmcm and \cmcii space group tend to have band gaps in the lower part of the spectra (about \SI{1}{\electronvolt}), whereas \piic alloys tend towards the upper region of the spectra (about \SI{2}{\electronvolt}). These compounds open up the \mixing space and can also be considered for future band gap tuning. Thus, we assume that at least a part of compositions within this \mixing space may have a direct band gap that fulfils our materials design criterion, whereby these alloys might also be of interest.

\clearpage
\pdfbookmark[section]{References}{references}
\bibliography{lit_a2bch2x3}